\documentclass[aps,prb,reprint,superscriptaddress]{revtex4-1}

\pdfoutput=1

\usepackage{graphicx,color}

\usepackage{amsmath,amssymb}
\usepackage{longtable}

\begin{document}


\title{
Vibronic coupling in C$_{60}^-$ anion revisited: Precise derivations 
from photoelectron spectra and DFT calculations}


\author{Naoya Iwahara}
\affiliation{Department of Molecular Engineering, Graduate School of Engineering,
Kyoto University, Kyoto 615-8510, Japan}
\author{Tohru Sato}%
\email[]{tsato@scl.kyoto-u.ac.jp}
\affiliation{Department of Molecular Engineering, Graduate School of Engineering,
Kyoto University, Kyoto 615-8510, Japan}
\affiliation{Fukui Institute for Fundamental Chemistry, Kyoto University,
Takano-Nishihiraki-cho 34-4, Sakyo-ku, Kyoto 606-8103, Japan}
\author{Kazuyoshi Tanaka}
\affiliation{Department of Molecular Engineering, Graduate School of Engineering,
Kyoto University, Kyoto 615-8510, Japan}
\author{Liviu F. Chibotaru}
\email[]{Liviu.Chibotaru@chem.kuleuven.be}
\affiliation{Division of Quantum and Physical Chemistry, University of Leuven, 
Celestijnenlaan 200F, B-3001 Leuven, Belgium}



\begin{abstract}
The vibronic coupling constants of C$_{60}^-$ are derived from the 
photoelectron spectrum measured by Wang {\it et al}. 
[X. B. Wang, H. K. Woo, and L. S. Wang, 
J. Chem. Phys., {\bf 123}, 051106 (2005).]
at low temperature with high-resolutions.
We find that the couplings of the Jahn--Teller modes of C$_{60}^-$ are 
weaker than the couplings reported by Gunnarsson {\it et al}. 
[O. Gunnarsson, H. Handschuh, P. S. Bechthold, B. Kessler,  
G. Gantef{\"{o}}r, and W. Eberhardt, Phys. Rev. Lett., {\bf 74}, 1875 (1995).].
The total stabilization energy after $h_g$ and $a_g$ modes is reduced
with respect to the previous derivation of Gunnarsson {\it et al}. 
by 30 \%.
The computed vibronic coupling constants using DFT with B3LYP functional
agree well with the new experimental constants,
so the discrepancy between theory and experiment persistent in the previous
studies is basically solved.
\end{abstract}


\maketitle

\section{INTRODUCTION
\label{SEC:INTRODUCTION}}

Much attention has been paid to the Jahn--Teller effect of fullerene 
(C$_{60}$) in various electronic states not only because the Jahn--Teller 
effect is an interesting problem in molecular physics\cite{Chancey1997a}
but also because it is expected to play an important role in the 
mechanism of the superconductivity in alkali-doped fullerides. 
\cite{Gunnarsson2004a}
Thus, the strength of the
electron-vibration coupling (vibronic coupling) of C$_{60}$ 
which characterizes the Jahn--Teller effect has
been one of the important topics.
The vibronic coupling constants (VCCs) 
have been estimated experimentally
{
\cite{Gunnarsson1995a, Winter1996a, Hands2008a}
}
and theoretically\cite{Varma1991a, Schluter1992a, Faulhaber1993a, 
Antropov1993a,  Breda1998a, Manini2001a, Saito2002a, Frederiksen2008a, 
Janssen2010a}.

In the experimental studies of vibronic coupling in fullerene, 
a landmark is the photoelectron spectroscopy (PES) of C$_{60}^-$
in gas phase by Gunnarsson {\it et al}.\cite{Gunnarsson1995a} 
As C$_{60}^-$ is one of the most studied systems,
in addition to this experimental work,
computational works have been performed by many authors.
However, discrepancy between the coupling constants of 
the experimental and theoretical works have been reported.
\cite{Gunnarsson1995a, Manini2001a, Saito2002a}
The theoretical stabilization energies as estimated by density
functional theory (DFT) calculation were always obtained smaller than 
that derived from the experiment of Gunnarsson {\it et al}.
Besides uncertainties intrinsic to the DFT method, whose predictions 
depend on the used exchange-correlation functional, 
one should note that the derivation of vibronic coupling constants in 
Ref.~\onlinecite{Gunnarsson1995a} is not perfect either.
First, the thermal excitations were not included in the simulation,
although the vibrational temperature of C$_{60}^-$ was estimated 
about 200 K in the experiment.
{
Second, not all vibronic coupling constants have been estimated from 
the spectrum because of the low resolution.
}
For the same reason, the computed VCCs of totally symmetric modes 
were used to simulate the PES.

Recently, Wang {\it et al}. remeasured photoelectron spectra of C$_{60}^-$.
\cite{Wang2005a}
In their experiment, the vibrational temperature of C$_{60}^-$ 
is between 70 K and 90 K and the resolution is about 16 meV,
i.e., much smaller than the resolution of 40 meV 
in the experiment of Gunnarsson {\it et al}.
Accordingly, the spectrum of Wang {\it et al}. is narrower and 
has more structures, 
therefore, it is expected 
to yield more reliable coupling constants.

In this work, we simulate the photoelectron spectra of Wang {\it et al}. 
\cite{Wang2005a} and Gunnarsson {\it et al}.\cite{Gunnarsson1995a} and
give new derivations of the VCCs of C$_{60}^-$.
We also compute the VCCs of C$_{60}^-$ using the DFT method
and compare them with the experimental values.

\section{THE SOLUTION OF THE JAHN--TELLER PROBLEM OF C$_{60}^-$
\label{SEC:JT}}
The equilibrium geometry of neutral fullerene is taken as the reference
nuclear configuration.
At this reference structure, 
the ground electronic state of C$_{60}^-$ is $T_{1u}$. 
According to the selection rule, the $T_{1u}$ electronic state couples 
with two $a_g$ and eight $h_g$ vibrational modes: 
\begin{eqnarray}
 \left[T_{1u}^2\right] = a_g \oplus h_g.
\label{Eq:selection}
\end{eqnarray}
We consider the linear $T_{1u} \otimes (2 a_g \oplus 8 h_g)$ 
Jahn--Teller Hamiltonian. The Hamiltonian is written as follows:
\begin{eqnarray}
 H &=& \sum_{i=1}^2 \left[\frac{1}{2}\left(
 P_{a_g(i)}^2 + \omega_{a_g(i)}^2 Q_{a_g(i)}^2 + V_{a_g(i)} Q_{a_g(i)}
\right)\right] \hat{I}
\nonumber \\
&+& \sum_{\mu=1}^8 \sum_{m=-2}^2 \left[\frac{1}{2}\left(
 P_{h_g(\mu)m}^2 + \omega_{h_g(\mu)}^2 Q_{h_g(\mu)m}^2
\right) \hat{I} 
\right.
\nonumber \\
&+&
\left.
  \sqrt{\frac{5}{2}} (-1)^m V_{h_g(\mu)} 
Q_{h_g(\mu)m} \hat{C}_{-m}
\right],
\label{Eq:JTH}
\end{eqnarray}
where $Q_{\Gamma(\mu)m}$ is the mass-weighted normal coordinate of 
$m$ element of the $\Gamma(\mu)$ mode ($\Gamma = a_g, h_g$), 
$P_{\Gamma(\mu)m}$ is the conjugate momentum of the normal coordinate
$Q_{\Gamma(\mu)m}$, $\omega_{\Gamma(\mu)}$ is the frequency of 
the $\Gamma(\mu)$ mode, $V_{\Gamma(\mu)}$ is the VCC of the $\Gamma(\mu)$ mode,
and $\hat{I}$ and $\hat{C}_{-m}$ are the $3 \times 3$ unit matrix
and a matrix whose elements are Clebsch--Gordan coefficients, respectively.
The normal modes and frequencies of C$_{60}$ are used for C$_{60}^-$ , 
so the higher vibronic terms which mix the normal modes of 
fullerene are neglected.
As a $T_{1u}$ electronic basis set 
$\{|m_{\rm el}\rangle ;m_{\rm el} = -1, 0, 1\}$
and normal coordinates of the $h_g$ modes 
$\{Q_{h_g(\mu)m} ;m = -2, -1, 0, 1, 2\}$, we use complex basis 
which transform as spherical harmonics 
$\{Y_{1m_{\rm el}}; m_{\rm el} = -1, 0, 1\}$ and 
$\{Y_{2m}; m = -2, -1, 0, 1, 2\}$, respectively, under the rotations.
\cite{OBrien1971a, Auerbach1994a}
Then $\hat{I}$ and $\hat{C}_{-m}$ are written as 
\cite{Edmonds1974a}
\begin{eqnarray}
 \hat{I} &=&
 \begin{pmatrix}
  1 & 0 & 0 \\
  0 & 1 & 0 \\
  0 & 0 & 1 
 \end{pmatrix}
,
\quad
 \hat{C}_{-2} =
 \begin{pmatrix}
  0 & 0 & \sqrt{\frac{3}{5}} \\
  0 & 0 & 0 \\
  0 & 0 & 0 
 \end{pmatrix}
,
\nonumber 
\\
 \hat{C}_{-1} &=& 
 \begin{pmatrix}
  0 & -\sqrt{\frac{3}{10}} & 0 \\
  0 & 0 & \sqrt{\frac{3}{10}} \\
  0 & 0 & 0 
 \end{pmatrix}
,
\quad
 \hat{C}_0 =
 \begin{pmatrix}
  \frac{1}{\sqrt{10}} & 0 & 0 \\
  0 & \frac{-2}{\sqrt{10}} & 0 \\
  0 & 0 & \frac{1}{\sqrt{10}}
 \end{pmatrix}
,
\nonumber \\
 \hat{C}_{1} &=&
 \begin{pmatrix}
  0 & 0 & 0 \\
  \sqrt{\frac{3}{10}} & 0 & 0 \\
  0 & -\sqrt{\frac{3}{10}} & 0 
 \end{pmatrix}
,
\quad
 \hat{C}_{2} = 
 \begin{pmatrix}
  0 & 0 & 0 \\
  0 & 0 & 0 \\
  \sqrt{\frac{3}{5}} & 0 & 0 
 \end{pmatrix}
.
\label{Eq:CG}
\end{eqnarray}
This type of the Jahn--Teller problem was investigated by O'Brien
\cite{OBrien1969a} and the vibronic coupling constants defined by her
are often used.
Thus we introduce the coefficient $\sqrt{5/2}$ in front of the vibronic 
term to make $V_{h_g}$ the same as O'Brien's coupling constants.

Since the linear $T_{1u} \otimes (2 a_g \oplus 8 h_g)$ Jahn--Teller 
Hamiltonian (\ref{Eq:JTH}) commutes with 
squared vibronic angular momentum $\mathbf{J}$ and the $z$ component
of $\mathbf{J}$,\cite{Bersuker1989a}
the eigenstate of the Hamiltonian (\ref{Eq:JTH}) is the simultaneous
eigenstate of the vibronic angular momentum $J$, the $z$ component
of the vibronic angular momentum $M$.
Here, the vibronic angular momentum $\mathbf{J}$ is the sum of the 
vibrational angular momentum $\mathbf{L}$ and the ``energy spin" $\mathbf{S}$
describing the threefold orbital degeneracy ($S=1$).\cite{Bersuker1989a}
In the case of linear vibronic coupling, the eigenstate of $H$ 
is the product of the $T_{1u} \otimes (8h_g)$ Jahn--Teller part 
and the $a_g$ vibrational part.
As a vibronic basis 
a set of the products of electronic states and 
vibrational states of the $a_g$ and $h_g$ modes is used:
\begin{eqnarray}
\left\{|m_{\rm el}\rangle |\cdots \mathbf{n}_\mu \cdots \rangle
{
|v_1 v_2 \rangle_{a_g}
}
\right\}.
\label{Eq:vibrobasis}
\end{eqnarray}
Here, $\mathbf{n}_\mu$ means a set of vibrational quantum numbers of 
the $h_g(\mu)$ mode, $\mathbf{n}_\mu = \left\{n_{\mu m}\right\}$,
{ $v_1, v_2$}
are vibrational quantum numbers of the $a_g(1)$ mode 
and the $a_g(2)$ mode respectively. 
Then the eigenstate 
{
$|\Psi_{v_1 v_2 kJM}\rangle$ 
}
of the Hamiltonian (\ref{Eq:JTH}) which belongs to the eigenvalue
{
$\sum_{i=1}^2 \left[\hslash \omega_{a_g(i)} v_i - 
V_{a_g(i)}^2/(2 \omega_{a_g(i)}^2)\right] + E_{kJ}$
}
is represented as a linear combination of the vibronic basis
with constants $C^{\rm JT}_{m_{\rm el}, \mathbf{n}_1\cdots \mathbf{n}_8;kJM}$.
\begin{eqnarray}
{
 |\Psi_{v_1 v_2 kJM}\rangle
} 
& = &
 \sum_{m_{\rm el} = -1}^1 \sum_{\mathbf{n}_1} \cdots \sum_{\mathbf{n}_8}
 |m_{\rm el}\rangle |\mathbf{n}_1 \cdots \mathbf{n}_8 \rangle
\nonumber 
\\
 &\times& C^{\rm JT}_{m_{\rm el}, \mathbf{n}_1 \cdots \mathbf{n}_8;kJM}
\nonumber 
\\
 & \times &
 \sum_{v'_1=0}^\infty \sum_{v'_2=0}^\infty
 |v'_1 v'_2 \rangle_{a_g}
 S_{v'_1 v_1}(g_{a_g(1)}) S_{v'_2 v_2}(g_{a_g(2)}),
\nonumber\\
\label{Eq:vibrostate}
\end{eqnarray}
where 
$E_{kJ}$ is an eigenvalue of the $T_{1u} \otimes (8h_g)$ Jahn--Teller
Hamiltonian,
$k$ distinguishes energy levels with the same $J$ and $M$,
the dimensionless VCC of the $\Gamma(\mu)$ 
mode $g_{\Gamma(\mu)}$ ($\Gamma = a_g, h_g$) is defined as
\begin{eqnarray}
 g_{\Gamma(\mu)} &=& 
\frac{V_{\Gamma(\mu)}}{\sqrt{\hslash \omega_{\Gamma(\mu)}^3}},
\label{Eq:dimlessVCC}
\end{eqnarray}
the Franck--Condon factor of the $a_g$ mode { $S_{v'v}(g)$} 
is written as
{ 
\begin{eqnarray}
 S_{v'v}(g) &=&
\sqrt{\frac{v'! v!}{2^{v' - v}}}
e^{-\frac{1}{4}g^2}
\sum_{l = l_{\rm min}}^v
\left(-\frac{1}{2}\right)^l
\frac{g^{2l + v' - v}}{l! (v - l)! (v' - v + l)!},
\nonumber\\
\label{Eq:coef_ag}
\end{eqnarray}
}
$l_{\rm min} = 0$ for { $v \le v'$}, 
and { $l_{\rm min} = v-v'$ for $v>v'$}.
The origin of the energy is the lowest energy of C$_{60}^-$
without vibronic couplings.

To obtain the vibronic states, we diagonalize the linear 
$T_{1u} \otimes 8 h_g$ Jahn--Teller Hamiltonian
numerically using Lanczos method.
We use a truncated vibronic basis set,
\begin{eqnarray}
\left\{|m_{\rm el}\rangle 
|\cdots \mathbf{n}_\mu \cdots \rangle 
 ;\sum_{\mu=1}^8 \sum_{m=-2}^2 n_{\mu m} \le N  \right\}.
\label{Eq:truncated_vibrobasis}
\end{eqnarray}
Here, $N$ is the maximum number of the vibrational excitations
in the vibronic basis set (\ref{Eq:truncated_vibrobasis}).
We treat the vibronic states which $J$s are from $0$ to $7$.
Frequencies $\omega_{a_g(i)}, \omega_{h_g(\mu)}$ 
are taken from the experimental frequencies of 
Raman scattering in solid state C$_{60}$.\cite{Bethune1991a} 

Lastly, we introduce stabilization energies
which we use to show our results. 
The stabilization energy of each mode is defined as 
\begin{eqnarray}
 E_{s,i} &=& \frac{V_{a_g(i)}^2}{2 \omega_{a_g(i)}^2}, \\
 E_{{\rm JT},\mu} &=& \frac{V_{h_g(\mu)}^2}{2 \omega_{h_g(\mu)}^2}, 
\label{Eq:partialEsEJT}
\end{eqnarray}
and the total stabilization energies of the $a_g$ modes and $h_g$ modes are
\begin{eqnarray}
 E_s &=& \sum_{i=1}^2 E_{s,i}, \\
 E_{\rm JT} &=& \sum_{\mu=1}^8 E_{{\rm JT},\mu}.
\label{Eq:EsEJT}
\end{eqnarray}
They represent the depth of the potential energy surface from the 
energy of undistorted fullerene monoanion.

\section{SIMULATION OF THE PHOTOELECTRON SPECTRUM
\label{SEC:PES}}
The photoelectron spectrum is simulated within the sudden approximation.
\cite{Hedin1969a}
We assume that each C$_{60}^-$ is in a thermal equilibrium state,
hence we use a Boltzmann's distribution to calculate the statistical weight.
With these assumptions, the intensity of the transition which appears
at the binding energy $\hslash \Omega$ is written as follows:
\begin{widetext}
\begin{eqnarray}
 I(\Omega) &\propto&
 \sum_{k,J} 
{
 \sum_{v'_1,v'_2} 
 p_{v'_1} p_{v'_2}
} 
p_{kJ}
 \sum_{m_{\rm el} = -1}^1 
 \left| C^{\rm JT}_{m_{\rm el}, \mathbf{n}_1 \cdots \mathbf{n}_8;kJ0}
{
 S_{v'_1 v_1}(g_{a_g(1)}) S_{v'_2 v_2}(g_{a_g(2)})
}
\right|^2
\nonumber
\\
&\times& \delta\left[\frac{E_0 { + E_s} - E_{kJ}}{\hslash} 
 + \sum_{\mu=1}^8 \sum_{m=-2}^2 \omega_{h_g(\mu)} n_{\mu m}
 + \sum_{i=1}^2 \omega_{a_g(i)}{ (v_i-v'_i)} - \Omega \right],
\label{Eq:Intensity}
\end{eqnarray}
\end{widetext}
where, { $p_{v_i}$} and $p_{kJ}$ are the statistical weights of 
the $a_{g(i)}$ mode and the Jahn--Teller part, respectively, 
\begin{eqnarray}
{ p_{v_i}} 
&=& \frac{1}{Z_{a_g}} 
 \exp\left(-\hslash \omega_{a_g(i)} { v_i} \beta \right), 
\\
 p_{kJ} &=&
 \frac{2J+1}{Z_{\rm JT}}\exp\left(- E_{kJ}\beta\right),
\label{Eq:StatWeight}
\end{eqnarray}
$Z_{a_g}$ and $Z_{\rm JT}$ are 
corresponding statistical sums, and $E_0$ is the gap between the 
ground electronic energies of C$_{60}$ and C$_{60}^-$.
The envelope function is represented by using the 
Gaussian function with the standard deviation $\sigma$:
\begin{eqnarray}
 F(\Omega) = \int_{-\infty}^\infty I(\Omega') 
 \exp\left[\frac{-(\Omega-\Omega')^2}{2\sigma^2}\right]
 d\Omega'.
\label{Eq:PES}
\end{eqnarray}
For a decent simulation of experimental PES one should include in 
Eq. (\ref{Eq:PES}), in principle, also the contributions from the rotational 
spectrum of C$_{60}^-$. 
However, due to a large momentum of inertia of fullerene and restrictive 
selection rules for the transitions between different rotational levels
\cite{Child1961a} our estimations gave an expected enlargement of the 
transition band of only several wave numbers. 
This is negligible compared the full width at half maximum
(FWHM) given by the envelope function
(\ref{Eq:PES}). 

To evaluate the agreement between the simulated spectrum and the experimental
spectrum, we calculate the residual of theoretical spectrum 
$F_{\rm calc}(\Omega)$ and the experimental spectrum $F_{\rm exp}(\Omega)$.
The residual $R$ is defined by the equation:
\begin{eqnarray}
R &=& \min_{f,\Omega_{\rm shift}} 
\left\{
\frac{
 \sum_{i=0}^M \left[F_{\rm calc}(\Omega_i) - f 
F_{\rm exp}(\Omega_i-\Omega_{\rm shift})\right]^2}{
\sum_{j=0}^M F_{\rm calc}^2(\Omega_j)}
\right\}.
\nonumber\\
\label{Eq:R}
\end{eqnarray}
Here, $f$ is the parameter to vary the height, 
$\Omega_{\rm shift}$ is the parameter to shift the experimental spectrum,
$\Omega_i$ is a sampling point. 
The minimum and maximum of $\Omega_i$ is $\Omega_{\rm min}$ and 
$\Omega_{\rm max}$, and the gap between adjacent sampling points 
$\Delta \Omega$ is constant.
Then $M$ is represented as $M=(\Omega_{\rm max} - \Omega_{\rm min})/
\Delta \Omega$ and $\Omega_i = \Omega_{\rm min} + i \Delta \Omega$.
In the calculation of the residual $R$, 
$\Omega_{\rm min}$, $\Omega_{\rm max}$, and $\Delta \Omega$ are
$-$200 cm$^{-1}$, 1600 cm$^{-1}$, and 0.5 cm$^{-1}$ respectively.
We avoid the truncation of the zero phonon line
of Gunnarsson {\it et al}.\cite{Gunnarsson1995a}
VCCs are varied in order to make $R$ as small as possible within the accuracy 
of the experiment.
The accuracy is determined from the range of the 
vibrational temperature of C$_{60}^-$ in the experiment of Wang {\it et al}.
\cite{Wang2005a}
In their experiment, the vibrational temperature is between 70 K and 90 K. 
Although the shapes of the simulated spectra at 70 K and 90 K are different 
from each other, we cannot distinguish them from the experiment of 
Wang {\it et al}.
In terms of the residual $R$, the difference between $R$ at 70 K 
and $R$ at 90 K is 
practically indistinguishable.

\section{DFT CALCULATION OF VIBRONIC COUPLING CONSTANTS
\label{SEC:DFTVCC}}
The linear vibronic coupling constant of the $a_g(i)$ mode is 
a diagonal matrix element of the first derivative of the electronic 
Hamiltonian with respect to the normal coordinate at the reference 
geometry. \cite{Bersuker1989a}
\begin{eqnarray}
V_{a_g(i)} &=&
 \langle \psi|
 \left(
 \frac{\partial H_{\rm el}(\mathbf{R})}{\partial Q_{a_g(i)}}
 \right)_{\mathbf{R}_0}
 |\psi \rangle,
\label{Eq:linear_VCC_ag_int}
\end{eqnarray}
where $\psi$ is the ground electronic state. 
By applying the Hellmann--Feynman theorem \cite{Feynman1939a}
to Eq. (\ref{Eq:linear_VCC_ag_int}) and then 
transforming it into the formula with the vibrational vector, 
we obtain
\begin{eqnarray}
V_{a_g(i)} &=&
 \left(\frac{\partial E(\mathbf{R})}{\partial Q_{a_g(i)}}
 \right)_{\mathbf{R}_0}
\label{Eq:linear_VCC_ag_grad}
\\
 &=& \sum_{A=1}^{60}
 \left(
 \frac{\partial E(\mathbf{R})}{\partial \mathbf{R}_A}
 \right)_{\mathbf{R}_0}
 \cdot
 \frac{\mathbf{u}_A^{a_g(i)}}{\sqrt{M}} .
\label{Eq:linear_VCC_ag_grad_calc}
\end{eqnarray}
Here, $A$ indicates a carbon atom in C$_{60}$,
$\mathbf{R}_A$ is the Cartesian coordinate of $A$, 
$\mathbf{R}$ is the set of all $\mathbf{R}_A$, 
$H_{\rm el}(\mathbf{R})$ is the electronic Hamiltonian 
at the structure $\mathbf{R}$, 
$\mathbf{R}_0$ is the reference nuclear configuration, 
$E(\mathbf{R})$ is the ground electronic energy
$\langle \psi| H_{\rm el}(\mathbf{R})  |\psi\rangle$,
$M$ is the mass of carbon atom,
$\mathbf{u}_A^{a_g(i)}$ is the vibrational vector of the $a_g(i)$ mode.
Similarly, absolute value of the coupling constant of the $h_g(\mu)$ 
mode is written as
\begin{eqnarray}
  V_{h_g(\mu)}
   &=& 
{
   \sqrt{
    \sum_{m=-2}^2 
     \left(
      \frac{\partial E(\mathbf{R})}{\partial Q_{h_g(\mu)m}}
     \right)_{\mathbf{R}_0}^2
   }
}
\label{Eq:linear_VCC_hg_grad}
\\
   &=& 
   \sqrt{
   \sum_{m=-2}^2 \left[
    \sum_{A=1}^{60}
    \left(\frac{\partial E(\mathbf{R})}{\partial \mathbf{R}_A}
    \right)_{\mathbf{R}_0}
    \cdot \frac{\mathbf{u}_A^{h_g(\mu)m}}{\sqrt{M}}
   \right]^2
   }.
\label{Eq:linear_VCC_hg_grad_calc}
\end{eqnarray}
The equilibrium geometry $\mathbf{R}_0$, the vibrational vectors 
$\mathbf{u}^{a_g(i)}$, $\mathbf{u}^{h_g(\mu)m}$, and 
the gradient of the electronic energy 
$\left(\partial E(\mathbf{R})/\partial \mathbf{R}_A
\right)_{\mathbf{R}_0}$, entering the Eqs. 
(\ref{Eq:linear_VCC_ag_grad_calc}) and (\ref{Eq:linear_VCC_hg_grad_calc}), 
are obtained from {\it ab initio} calculations.
{
Note that the vibronic coupling constants
(\ref{Eq:linear_VCC_ag_grad}), (\ref{Eq:linear_VCC_hg_grad})
are not equal to the gradients of the frontier levels (see Appendix).
}

We compute the VCCs of C$_{60}^-$ using the DFT method. 
As 
exchange-correlation functional,
the hybrid functional of Becke
\cite{Becke1993a} (B3LYP) is used.
To find VCCs which are close to the experimental results the fraction of the 
Hartree--Fock exchange energy are varied from the original 
fraction 20\% to 30\% by 5\%.
We use the triple zeta basis sets, 6-311G(d), 6-311+G(d), and cc-pVTZ.

The structure optimization and the calculation of the vibrational modes 
are performed for the neutral fullerene.
The electronic wavefunction of C$_{60}^-$ are obtained from the variational 
calculation of an unrestricted Slater determinant.
As far as the method based on the single determinant is used, 
the spatial symmetry of the wavefunction is 
broken and the degeneracy of the singly occupied degenerate level is lifted.
\cite{Sato2006a, Sato2009a}
However, in the case of the cyclopentadienyl radical, 
it was demonstrated that the splitting of the total electronic energies
estimated by the unrestricted B3LYP method is only 0.4 meV.
\cite{Sato2006a, Sato2009a}
It is expected that the splitting of the $T_{1u}$ ground electronic energies
of C$_{60}^-$ is tiny and the symmetry of the electronic state
may not be broken significantly.
Thus we treat the wavefunction as a $T_{1u}$ wavefunction.
We calculate the energy gradient 
$\left.\partial E(\mathbf{R})/\partial \mathbf{R}_A
\right|_{\mathbf{R}\rightarrow \mathbf{R}_0}$ 
with the coupled perturbed Kohn--Sham method. 
In the calculation of the dimensionless VCCs (\ref{Eq:dimlessVCC})
and the stabilization energies (\ref{Eq:partialEsEJT}), 
we use the experimental frequencies.\cite{Bethune1991a}
To compute electronic structures we use the Gaussian 03 program.\cite{g03}

\section{Derivation of the vibronic coupling constants of $a_g$ modes from 
the structures of C$_{60}$ and C$_{60}^-$
\label{SEC:C-C}}
We also derive the stabilization energies of the $a_g$ modes 
from the experimental bond lengths of C$_{60}$ and C$_{60}^-$.
The structures of C$_{60}$ and C$_{60}^-$ with $I_h$ symmetry are determined 
by the C-C bond lengths of the edges between two hexagons (6:6) and
a hexagon and pentagon (6:5).
We use average 6:6 and 6:5 C-C bond lengths of TDAE-C$_{60}$ for C$_{60}^-$ 
and fullerite for C$_{60}$.
The data of TDAE-C$_{60}$ are obtained from the results of X-ray 
diffraction at 7K by Narymbetov {\it et al}. 
\cite{Narymbetov1999a} and at 25K and 90K by 
Fujiwara {\it et al}.\cite{Fujiwara2005a}.
The average bond lengths of fullerite are taken from the results of neutron 
diffraction at 5K by David {\it et al}.\cite{David1991a} 
and X-ray diffraction at 110K
by B\"{u}rgi {\it et al}.\cite{Burgi1992a}
To remove the thermal expansion of the C-C bond lengths, 
we use sets of bond lengths of C$_{60}$ and C$_{60}^-$ which are 
measured at close temperature.
That is, the bond lengths of C$_{60}$ measured at 5K is used with 
the bond lengths of C$_{60}^-$ measured at 7K and 25K, and the bond lengths 
of C$_{60}$ at 110K is used with the bond lengths of C$_{60}^-$ at 90K.
The vibronic coupling constants of the $a_g$ modes $V_{a_g(i)} (i=1,2)$ 
are obtained from the equation
\begin{align}
 V_{a_g(i)} = -\sum_{A=1}^{60} 
\left(\mathbf{R}_A - \mathbf{R}_{0,A}\right)\cdot 
\frac{\mathbf{u}_A^{a_g(i)}}{\sqrt{M}}.
\label{Eq:V_C-C}
\end{align}
To perform the calculation, we use the vibrational vector 
defined in the calculations with the B3LYP method
and the cc-pVTZ basis.

\section{RESULTS AND DISCUSSIONS
\label{SEC:RESULTS_DISCUSSIONS}}
\subsection{Simulation of the PES of Wang {\it et al}.
\label{SUBSEC:CALC_PES_WANG}}
\begin{figure}
 \includegraphics[height=8.5cm, angle=-90]{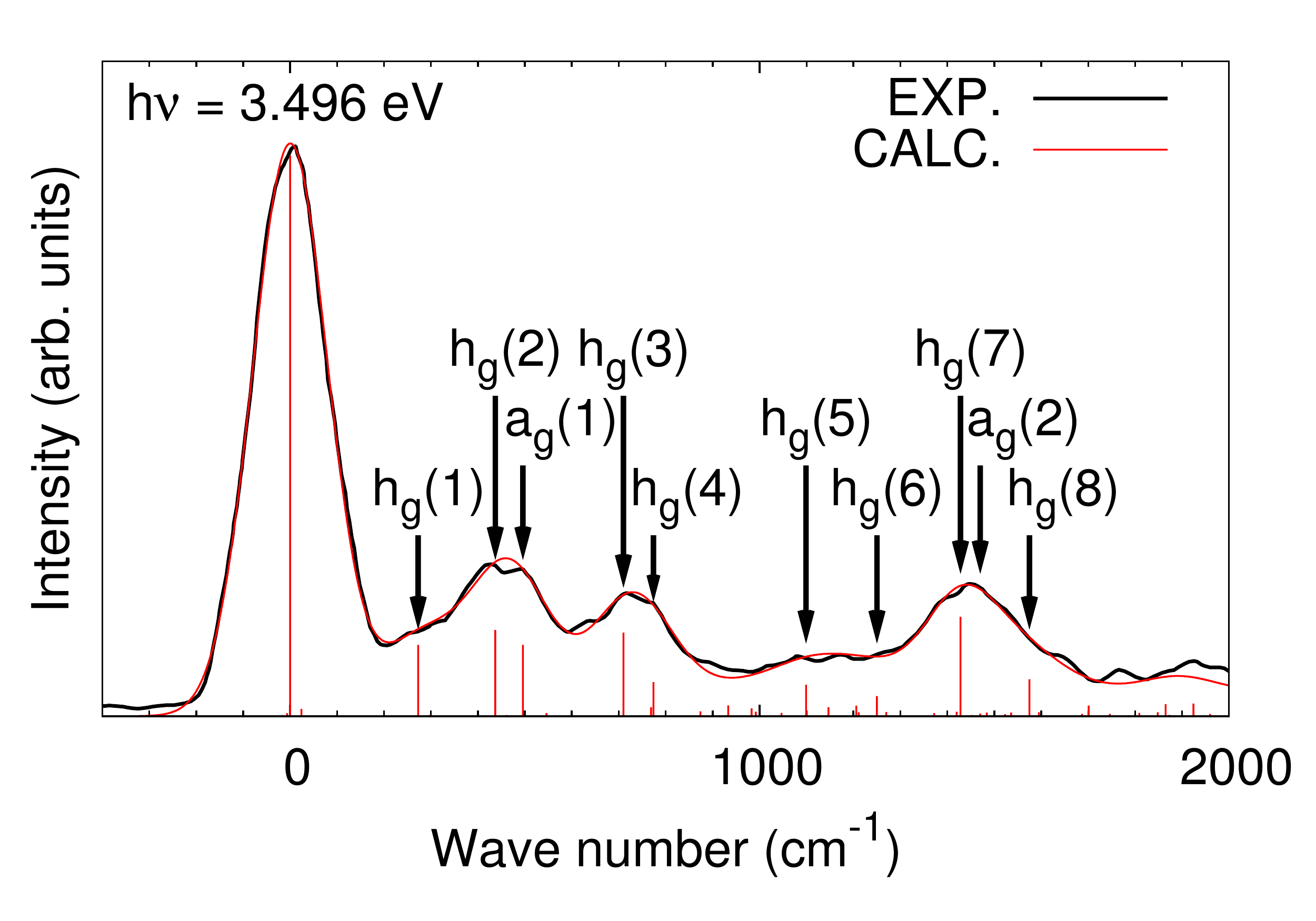}
 \caption{
The photoelectron spectrum measured by Wang {\it et al}. (black line)
and the simulated spectrum (red line). 
The simulation is performed at 70 K with $\sigma = 80$ cm$^{-1}$.
\label{FIG:PESWANG}}
\end{figure}
We simulate the photoelectron spectrum measured by Wang {\it et al}. 
\cite{Wang2005a} at 70 K. 
The basis set in Eq. (\ref{Eq:vibrostate})
includes up to 6 vibrational excitations ($N=6$).
The experimental and simulated spectra are shown in Fig. \ref{FIG:PESWANG}.
The transition between the ground states of C$_{60}^-$ and C$_{60}$
(the 0--0 line) \cite{0-0} is chosen as the origin of these spectra.
From the spectrum of Wang {\it et al}., we obtain several sets of VCCs
listed as (1), (2), and (3) in Tables \ref{TABLE:VCC} 
and \ref{TABLE:PRESENT_STABILIZATION}.
{
We extracted $\sigma = 80$ cm$^{-1}$ by fitting the FWHM of the 0--0 line
(188 cm$^{-1}$). 
The increase or decrease of the $\sigma$ makes the agreement
between the simulated and experimental spectra worse.
}
\begin{table*}
\caption{Absolute values of dimensionless vibronic coupling constants
obtained in the present work.
\label{TABLE:VCC}}
\begin{ruledtabular}
\begin{tabular}{ccccccccccccc}
     & Frequency & \multicolumn{5}{c}{PES}
     & \multicolumn{6}{c}{B3LYP\footnotemark[3]} \\
     &(cm$^{-1}$)& \multicolumn{3}{c}{Wang\footnotemark[1]} 
     & \multicolumn{2}{c}{Gunnarsson\footnotemark[2]} 
     & \multicolumn{3}{c}{6-311G(d)} & 
6-311+G(d) & \multicolumn{2}{c}{cc-pVTZ} \\
     &           &       &       &       &       &       & 20 \% & 25 \% & 30 \% & 20 \% & 20 \% & 25 \%\\
     &           &  (1)  &  (2)  &  (3)  &  (4)  &  (5)  &  (6)  &  (7)  &  (8)  &  (9)  &  (10) &  (11) \\
 \hline
 $a_g(1)$ &  496 & 0.505 & 0.505 & 0.500 & 0.141 & 0.505 & 0.287 & 0.272 & 0.269 & 0.346 & 0.289 & 0.286 \\
 $a_g(2)$ & 1470 & 0.100 & 0.200 & 0.300 & 0.424 & 0.200 & 0.415 & 0.445 & 0.460 & 0.455 & 0.430 & 0.450 \\
 $h_g(1)$ &  273 & 0.500 & 0.500 & 0.490 & 0.740 & 0.820 & 0.436 & 0.437 & 0.444 & 0.426 & 0.442 & 0.452 \\
 $h_g(2)$ &  437 & 0.525 & 0.520 & 0.515 & 0.860 & 0.690 & 0.498 & 0.504 & 0.508 & 0.479 & 0.498 & 0.494 \\
 $h_g(3)$ &  710 & 0.465 & 0.460 & 0.455 & 0.390 & 0.350 & 0.418 & 0.464 & 0.476 & 0.412 & 0.403 & 0.414 \\
 $h_g(4)$ &  774 & 0.310 & 0.310 & 0.300 & 0.490 & 0.490 & 0.259 & 0.241 & 0.243 & 0.252 & 0.273 & 0.283 \\
 $h_g(5)$ & 1099 & 0.285 & 0.280 & 0.280 & 0.320 & 0.300 & 0.211 & 0.233 & 0.241 & 0.211 & 0.212 & 0.217 \\
 $h_g(6)$ & 1250 & 0.220 & 0.230 & 0.235 & 0.190 & 0.160 & 0.126 & 0.169 & 0.178 & 0.126 & 0.125 & 0.124 \\
 $h_g(7)$ & 1428 & 0.490 & 0.470 & 0.435 & 0.320 & 0.430 & 0.398 & 0.414 & 0.433 & 0.392 & 0.398 & 0.415 \\
 $h_g(8)$ & 1575 & 0.295 & 0.285 & 0.260 & 0.350 & 0.410 & 0.338 & 0.335 & 0.345 & 0.330 & 0.333 & 0.343 \\
\end{tabular}
\end{ruledtabular}
\footnotetext[1]{
(1),(2),and (3) are derived from the PES of Wang {\it et al}.
(Ref.~\onlinecite{Wang2005a})}
\footnotetext[2]{
(4),(5) are derived from the PES of Gunnarsson {\it et al}.
(Ref.~\onlinecite{Gunnarsson1995a}) 
}
\footnotetext[3]{
The percentage 20 \%, 25 \%,  and 30 \% indicate fractions of 
the Hartree--Fock exact exchange taken in the exchange-correlation functional.
}
\end{table*}
\begin{table*}
\caption{Stabilization energies (meV) obtained in the present work.
\label{TABLE:PRESENT_STABILIZATION}}
\begin{ruledtabular}
\begin{tabular}{cccccccccccccccc}
& Frequency & \multicolumn{5}{c}{PES} & \multicolumn{6}{c}{B3LYP\footnotemark[3]}\\
& (cm$^{-1}$) & \multicolumn{3}{c}{Wang\footnotemark[1]} & 
\multicolumn{2}{c}{Gunnarsson\footnotemark[2]} & \multicolumn{3}{c}{6-311G(d)} 
& 6-311+G(d) & \multicolumn{2}{c}{cc-pVTZ}\\
     &           &       &       &       &       &       & 20 \% & 25 \% & 30 \% & 20 \% & 20 \% & 25 \%\\
     &           &  (1)  &  (2)  &  (3)  &  (4)  &  (5)  &  (6)  &  (7)  &  (8)  &  (9)  &  (10) &  (11) \\
\hline
 $a_g(1)$ &  496 & 7.8 & 7.8 & 7.7 & 0.6 & 7.8 & 2.5 & 2.3 & 2.2 & 3.7 & 2.6 & 2.5 \\
 $a_g(2)$ & 1470 & 0.9 & 3.6 & 8.2 &16.4 & 3.6 &15.7 &18.0 &19.3 &18.9 &16.9 &18.5 \\
 $h_g(1)$ &  273 & 4.2 & 4.2 & 4.1 & 9.3 &11.4 & 3.2 & 3.2 & 3.3 & 3.1 & 3.3 & 3.5 \\
 $h_g(2)$ &  437 & 7.5 & 7.3 & 7.2 &20.0 &12.9 & 6.7 & 6.9 & 7.0 & 6.2 & 6.7 & 6.6 \\
 $h_g(3)$ &  710 & 9.5 & 9.3 & 9.1 & 6.7 & 5.4 & 7.7 & 9.5 &10.0 & 7.5 & 7.2 & 7.6 \\
 $h_g(4)$ &  774 & 4.6 & 4.6 & 4.3 &11.5 &11.5 & 3.2 & 2.8 & 2.8 & 3.1 & 3.6 & 3.8 \\
 $h_g(5)$ & 1099 & 5.5 & 5.3 & 5.3 & 7.0 & 6.1 & 3.0 & 3.7 & 4.0 & 3.0 & 3.0 & 3.2 \\
 $h_g(6)$ & 1250 & 3.9 & 4.1 & 4.3 & 2.8 & 2.0 & 1.2 & 2.2 & 2.5 & 1.2 & 1.2 & 1.2 \\
 $h_g(7)$ & 1428 &21.3 &19.6 &16.8 & 9.1 &16.4 &14.0 &15.2 &16.6 &13.6 &14.0 &15.2 \\
 $h_g(8)$ & 1575 & 8.5 & 7.9 & 6.6 &12.0 &16.4 &11.2 &11.0 &11.6 &10.7 &10.9 &11.5 \\
 \hline
 $E_s$    &      & 8.7 &11.4 &15.9 &17.0 &11.4 &18.2 &20.3 &21.5 &22.6 &19.5 & 21.0 \\
 $E_{\rm JT}$ &  &65.0 &62.3 &57.7 &78.4 &82.1 &50.2 &54.4 &57.8 &48.4 &49.9 & 52.6 \\ 
$E_s+E_{\rm JT}$&&73.7 &73.7 &73.6 &95.4 &93.5 &68.4 &74.7 &79.3 &71.0 &69.4 & 73.6 \\ 
\end{tabular}
\end{ruledtabular}
\footnotetext[1]{
(1),(2),and (3) are derived from the PES of Wang {\it et al}.
(Ref.~\onlinecite{Wang2005a})}
\footnotetext[2]{
(4),(5) are derived from the PES of Gunnarsson {\it et al}.
(Ref.~\onlinecite{Gunnarsson1995a})
}
\footnotetext[3]{
The percentage 20 \%, 25 \%,  and 30 \% indicate fractions of 
the Hartree--Fock exact exchange taken in the exchange-correlation functional.
}
\end{table*}

To assess the thermal population of the excited vibronic states,  
we calculate statistical weights 
of the excited Jahn--Teller levels $p_{kJ}$ at 70 K and 90 K.
The vibronic levels are obtained using the set of VCCs (1). 
In the calculation of the distribution function $Z_{\rm JT}$, 
we include all excited vibronic levels { whose} weights
are larger than $10^{-7}$.
The computed weights are shown in Table \ref{TABLE:WEIGHT-WANG}.
Although these statistical weights are computed using the set (1), 
rest of the sets of VCCs (2), (3) give similar results.
The statistical weights of the ground vibronic 
level at 70 K and 90 K are more than 90 \%.
This indicates that the transition from the ground vibronic 
level is dominant in the PES of Wang {\it et al}.
We focus, therefore, on the ground vibronic level to discuss 
the effect of the size of the basis (\ref{Eq:truncated_vibrobasis})
on the calculated vibronic states.
The ground vibronic level is $-962.65$ cm$^{-1}$ when we use the basis set 
with $N = 5$.
Compared with the gap between the ground and first excited vibronic levels 
with $J=1$, the change of the ground vibronic level due to the increase of 
the size of the vibronic basis set is only about 0.07 \%. 
Therefore, we regard our basis set as large enough to simulate the 
spectrum of Wang {\it et al}.
\begin{table}
\caption{The lowest vibronic levels (cm$^{-1}$) and the 
statistical weights $p_{kJ}$ (\%) at 70 K and 90 K.
$J$ is the magnitude of the vibronic angular momentum.
To calculate the vibronic levels, the set of VCCs (1) in Table 
\ref{TABLE:VCC} is used.
\label{TABLE:WEIGHT-WANG}}
\begin{ruledtabular}
 \begin{tabular}{ccccc}
   Level & $J$ &   Energy  & \multicolumn{2}{c}{Weight} \\
   & & &  70 K & 90 K \\
  \hline
    1 & 1 & -962.85 & 97.75 & 92.48 \\
    2 & 3 & -713.97 &  1.37 &  4.04 \\
    3 & 2 & -683.40 &  0.52 &  1.77 \\
    4 & 1 & -672.75 &  0.25 &  0.90 \\
  \hline
   Sum & & & 99.89 & 99.19 \\
 \end{tabular}
\end{ruledtabular}
\end{table}

\begin{figure}
 \includegraphics[height=5.5cm, angle=-90]{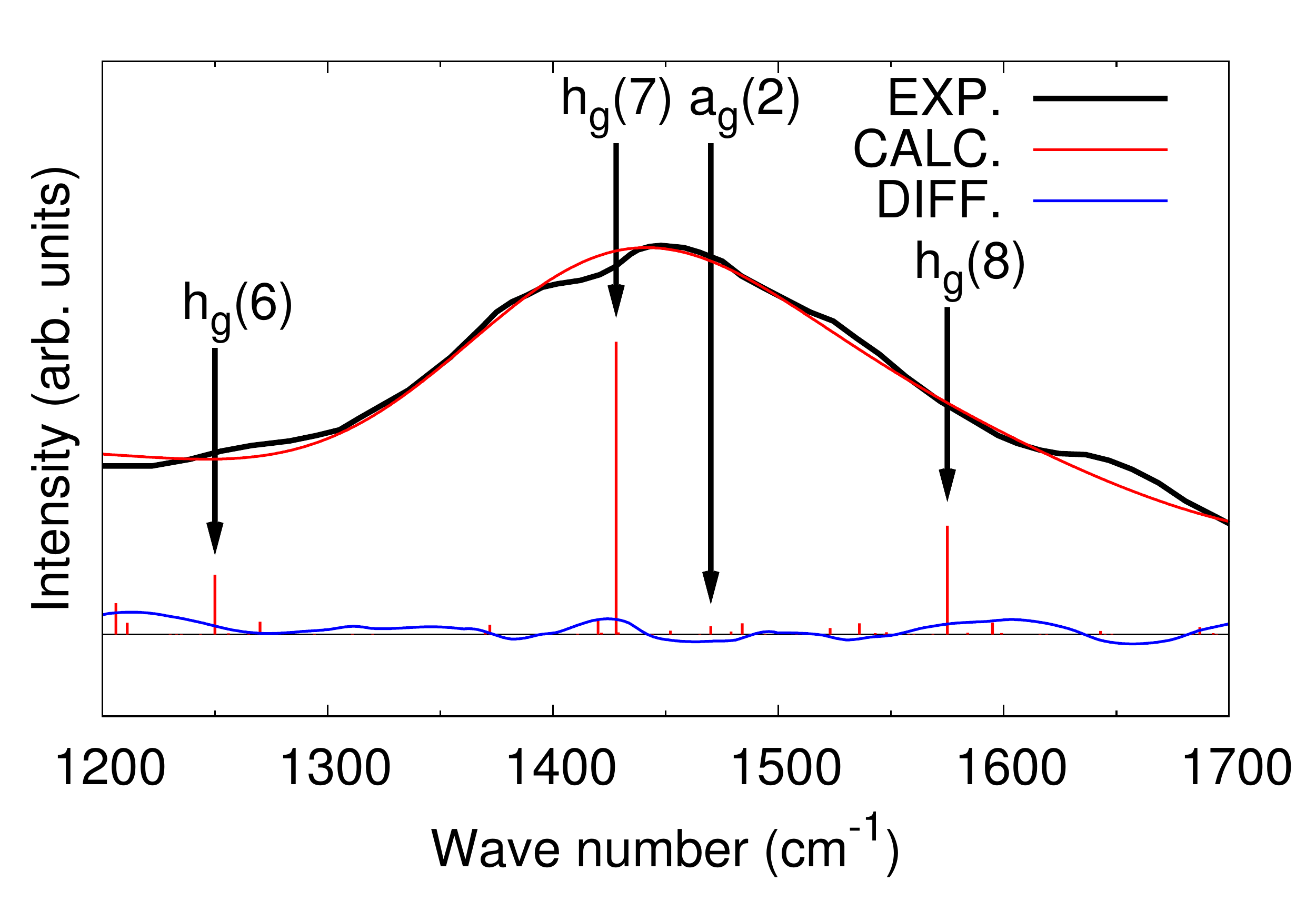}
\\
 \includegraphics[height=5.5cm, angle=-90]{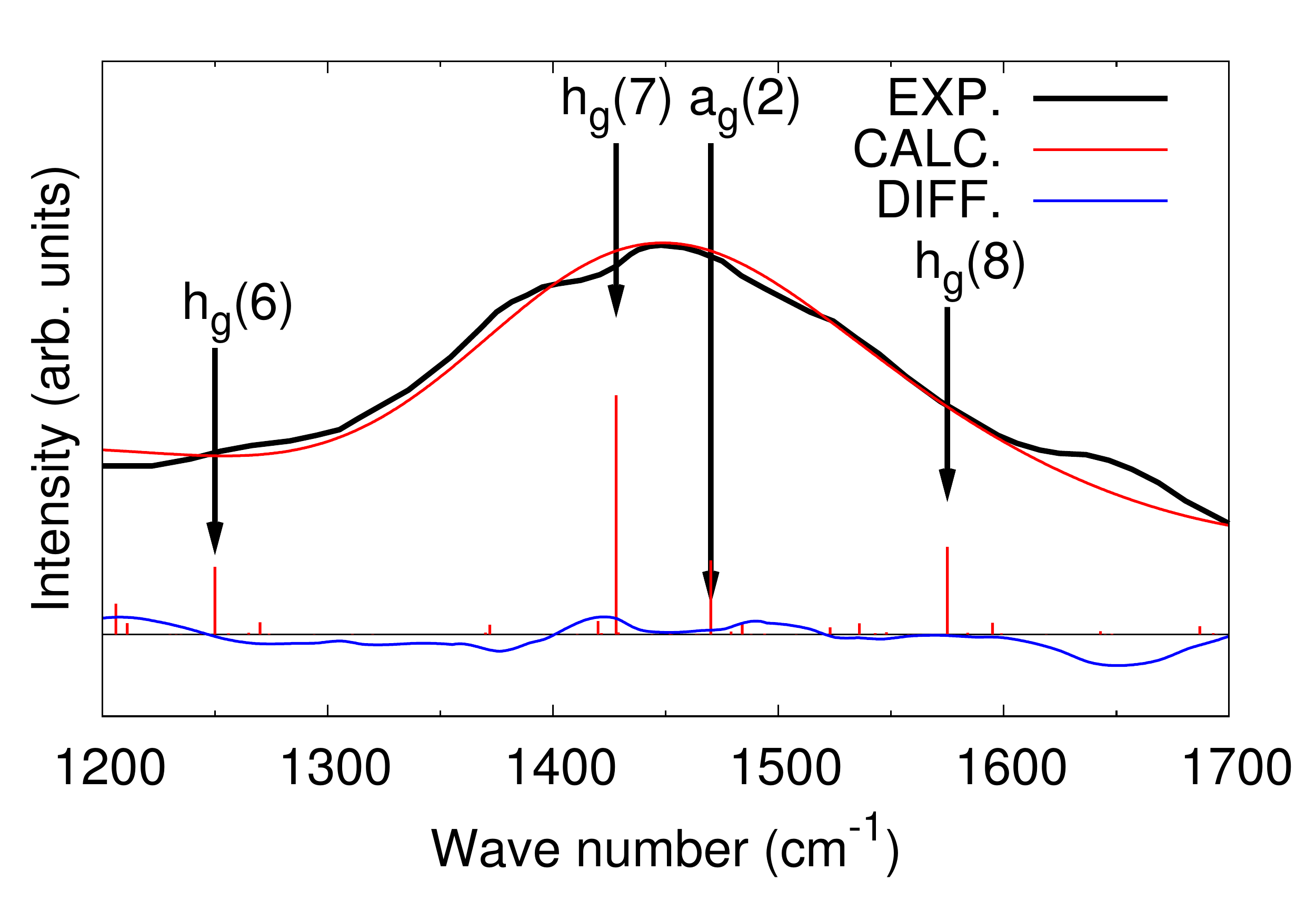}
 \caption{
The peak of the photoelectron spectra due to $a_g(2)$, $h_g(7)$, and 
$h_g(8)$ modes. 
The black line indicates the experimental spectrum of Wang {\it et al.}, 
\cite{Wang2005a}
the red line indicates the simulated spectrum, and the blue line indicates
the difference between the experimental and simulated spectra.
The top of the two spectra is simulated using the VCCs (1) and the bottom one
is simulated using the VCCs (3) from Table \ref{TABLE:VCC}.
The simulation is performed at 70 K with $N=6$ and $\sigma = 80$ cm$^{-1}$.
\label{FIG:PESWANG-DETAIL}}
\end{figure}

The differences between several sets of VCCs are in the constants of $a_g(2)$, 
$h_g(7)$, and $h_g(8)$ modes.
If we increase the dimensionless VCC (the stabilization energy) of the 
$a_g(2)$ mode from 0.1 to 0.3 (0.9 to 8.2 meV)
and at the same time decrease the dimensionless VCCs of $h_g(7), h_g(8)$ modes, 
the shape of the PES does not vary significantly 
(see Fig. \ref{FIG:PESWANG-DETAIL}).
This is due to a poor resolution of the peaks of $a_g(2), h_g(7)$, 
and $h_g(8)$ modes and essentially the same problem arose in the analysis 
of Gunnarsson {\it et al}.\cite{Gunnarsson1995a}
In the latter case, the stabilization energy of $a_g(2)$ is varied 
from 0 to 45 meV, i.e., in a range larger than ours. 
Owing to the narrow peaks of the spectrum of Wang {\it et al}.,
we can derive the VCCs with less ambiguity.

We compute the residuals (\ref{Eq:R}) of the experimental and the 
simulated spectra for all sets of VCCs at 70 K and 90 K. 
The values are shown in Table \ref{TABLE:RESIDUE_WANG}.
\begin{table}
\caption{Residuals of the experimental and simulated spectra.
The calculation of the residual is performed for all sets of VCCs 
in Table \ref{TABLE:VCC} at 70 K and 90 K.
\label{TABLE:RESIDUE_WANG}
}
 \begin{ruledtabular}
 \begin{tabular}{ccccccc}
  & \multicolumn{2}{c}{Set (1)} 
  & \multicolumn{2}{c}{Set (2)}  
  & \multicolumn{2}{c}{Set (3)} \\
  & 70 K & 90 K & 70 K & 90 K & 70 K & 90 K\\
  \hline
  $R \times 10^{-4}$ & 8.35 & 8.07 & 8.41 & 8.16 & 8.76 & 8.41\\
 \end{tabular}
 \end{ruledtabular}
\end{table}
The differences of the residuals of the different sets of VCCs are 
almost within the ambiguity of the vibrational temperature.
Therefore, we conclude that these sets of VCCs cannot be distinguished from 
the experiment of Wang {\it et al}.

Although we obtain several sets of VCCs, the stabilization 
energies $E_s+E_{\rm JT}$ are similar to each other (see (1), (2), and (3) in 
Table \ref{TABLE:PRESENT_STABILIZATION}).
On the other hand, present stabilization energies are smaller than 
the stabilization energy of Gunnarsson {\it et al}. \cite{Gunnarsson1995a}
by 30 \% (see Table \ref{TABLE:STABILIZATION}),
i.e., the Jahn--Teller coupling is weaker than previously expected.
We find that the distributions of $E_{s,i}$ and $E_{{\rm JT},\mu}$ also 
differ from each other (Tables \ref{TABLE:PRESENT_STABILIZATION}, 
\ref{TABLE:STABILIZATION}).
In Ref.~\onlinecite{Gunnarsson1995a}, the stabilization energy of $h_g(2)$
was found the strongest,
while our results show that the strongest is the 
stabilization energy of $h_g(7)$.

{
Hands {\it et al}. estimated the Jahn--Teller stabilization energy $E_{\rm JT}$ 
of 57.94 meV within the single-mode $T_{1u} \otimes h_g$ Jahn--Teller model 
from the visible and near-infrared spectrum. \cite{Hands2008a}
The present Jahn--Teller stabilization energy agrees well with their value.
}

\begin{table*}
\caption{
Comparison between the obtained stabilization energies 
of C$_{60}^-$ with previous results (meV).
\label{TABLE:STABILIZATION}}
\begin{ruledtabular}
\begin{tabular}{ccccccccccc}
& Freq. & \multicolumn{2}{c}{PES} & \multicolumn{2}{c}{LDA} & 
GGA & MNDO & \multicolumn{3}{c}{B3LYP}\\
& (cm$^{-1}$) & (3)\footnotemark[1] 
& Gun \cite{Gunnarsson1995a} 
& Man \cite{Manini2001a}
& Bre \cite{Breda1998a}
& Fre\cite{Frederiksen2008a}
& Var \cite{Varma1991a} \footnotemark[2] 
& Sai \cite{Saito2002a} \footnotemark[2] 
& { Laf \cite{Janssen2010a}}
& (10)\footnotemark[1] \\
\hline
 $a_g(1)$ &  496 & 7.7 & 0.6 & 0.2 & 1.5 & 1.5 &  -  &  -  &{ 1.8} & 2.6 \\
 $a_g(2)$ & 1470 & 8.2 &16.4 & 2.7 & 9.0 &11.0 &  -  &  -  &{16.4} &16.9 \\
 $h_g(1)$ &  273 & 4.1 &11.4 & 2.7 & 6.0 & 2.8 & 1.8 & 3.6 &{ 3.5} & 3.3 \\
 $h_g(2)$ &  437 & 7.2 &24.0 & 6.3 &15.6 & 7.0 & 0.6 & 6.6 &{ 6.5} & 6.7\\
 $h_g(3)$ &  710 & 9.1 & 7.8 & 5.5 & 6.6 & 6.1 & 0.6 & 6.6 &{ 7.1} & 7.2\\
 $h_g(4)$ &  774 & 4.3 &10.8 & 2.4 & 3.0 & 2.4 & 0.0 & 3.0 &{ 3.1} & 3.6\\
 $h_g(5)$ & 1099 & 5.3 & 7.2 & 2.6 & 3.6 & 2.6 & 3.6 & 3.0 &{ 3.0} & 3.0\\
 $h_g(6)$ & 1250 & 4.3 & 3.0 & 1.5 & 1.8 & 1.9 & 0.0 & 1.8 &{ 1.3} & 1.2\\
 $h_g(7)$ & 1428 &16.8 &10.2 & 9.0 & 9.6 & 9.0 &20.4 &13.2 &{13.8} &14.0\\
 $h_g(8)$ & 1575 & 6.6 &13.8 & 8.2 & 4.8 & 8.8 & 6.6 &10.2 &{10.6} &10.9 \\
 \hline
 $E_s$    &      &15.9 &17.0 & 2.9 &10.5 &12.5 &  -  &  -  &{18.2} &19.5 \\
 $E_{\rm JT}$ &  &57.7 &88.2 &38.2 &51.0 &40.6 &33.6 &48.0 &{48.9} &49.9 \\ 
$E_s+E_{\rm JT}$&&73.6 &105.2&41.1 &61.5 &53.1 &  -  &  -  &{67.1} &69.4 \\ 
\end{tabular}
\end{ruledtabular}
\footnotetext[1]{Results given in Tables \ref{TABLE:PRESENT_STABILIZATION}.}
\footnotetext[2]{The stabilization energies of the $a_g$ modes are 
not reported.}
\end{table*}

\subsection{Simulation of the PES of Gunnarsson {\it et al}.
\label{SUBSEC:CALC_PES_GUNNARSSON}}
\begin{figure}
 \includegraphics[height=8.5cm, angle=-90]{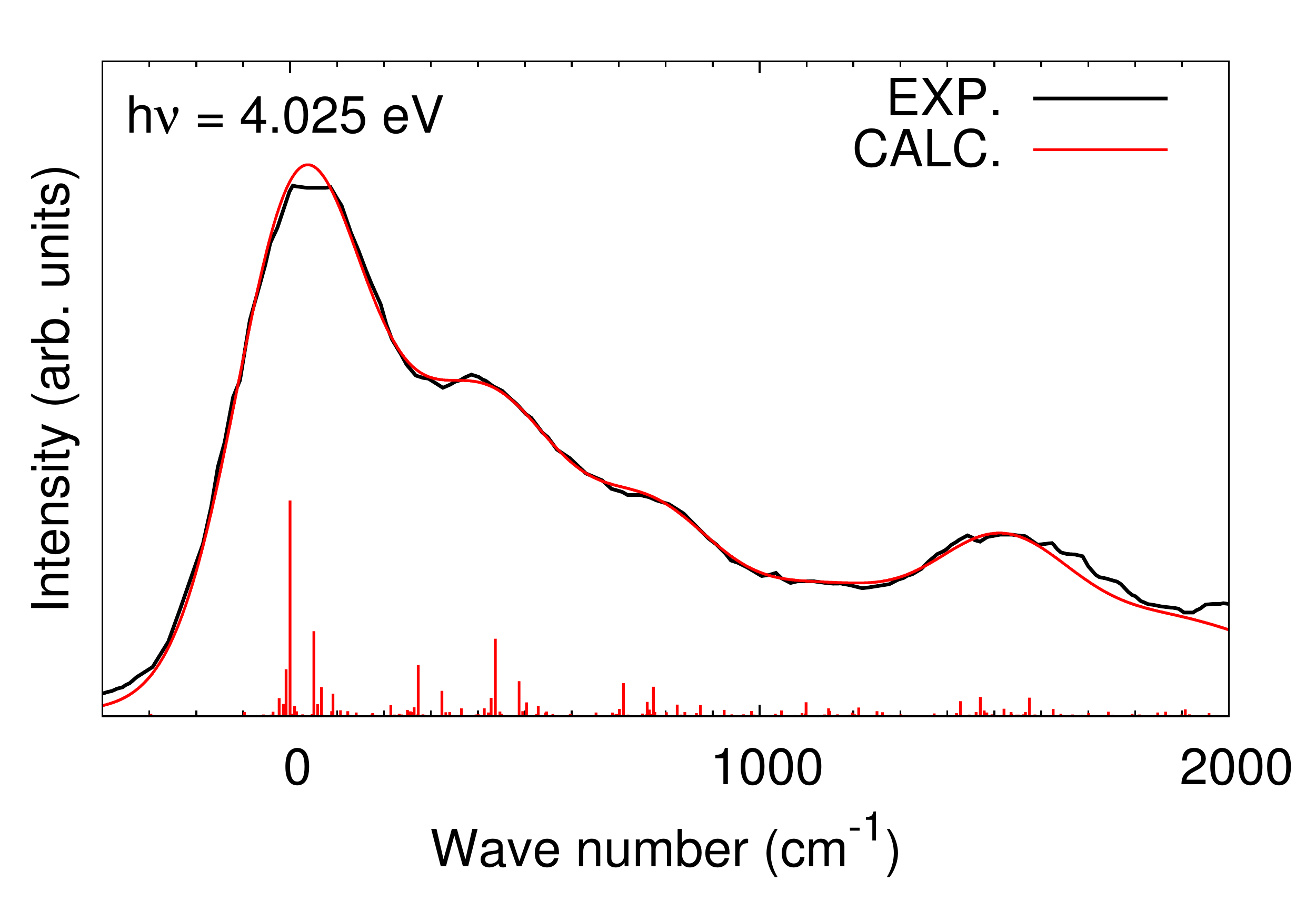}
 \caption{
The experimental photoelectron spectrum measured by Gunnarsson {\it et al.} 
\cite{Gunnarsson1995a}
(black line) and the simulated spectrum (red line). 
The simulation is performed at 200 K with $\sigma = 120$ cm$^{-1}$.
\label{FIG:PES_GUNNARSSON}}
\end{figure}
As a preliminary calculation, we compute the vibronic levels 
using the data from Ref.~\onlinecite{Gunnarsson1995a},
that is, the same VCCs and the same size of the vibronic basis ($N=5$). 
The statistical weight of the ground state at 200 K is obtained ca. 39 \%. 
This result indicates that not only the ground level but also excited 
levels must be considered in order to simulate the spectrum of 
Gunnarsson {\it et al}. \cite{Gunnarsson1995a}

We simulate this spectrum at 200 K with the FWHM of 283 cm$^{-1}$ 
($\sigma = 120$ cm$^{-1}$).
The size of the vibronic basis set is $N=7$.
The experimental and the simulated spectra are shown in 
Fig. \ref{FIG:PES_GUNNARSSON}.
As was mentioned also by Gunnarsson {\it et al}.,\cite{Gunnarsson1995a} 
we obtain several sets of VCCs that give 
close stabilization energies. 
These sets of dimensionless VCCs and stabilization energies are (4), (5) in 
Table \ref{TABLE:VCC} and \ref{TABLE:PRESENT_STABILIZATION} respectively.
In comparison with the original stabilization energy of Gunnarsson {\it et al}., 
present stabilization energies are smaller by 10 meV.
However, stabilization energies are still larger than those obtained from 
the spectrum of Wang {\it et al}. by 20 meV.
The inconsistencies of these VCCs come from the 
difference between the shapes of the spectra of Wang {\it et al}.
and Gunnarsson {\it et al}., due to different vibrational temperature
and resolution.

\begin{table}
\caption{
The computed vibronic levels (cm$^{-1}$) and statistical weights 
$p_{kJ}$ (\%) using the sets of VCCs (4), (5) which are 
derived from the experimental spectrum of Gunnarsson {\it et al}. 
\cite{Gunnarsson1995a}
The statistical weights are calculated at 200K. 
\label{TABLE:VIBRO_WEIGHT_GUNNARSSON}}
\begin{ruledtabular}
 \begin{tabular}{ccccccccc}
  Level & \multicolumn{3}{c}{Set (4)} & \multicolumn{3}{c}{Set (5)} \\
        & $J$ & Energy & Weight
        & $J$ & Energy & Weight\\
        & & & 200 K &  & & 200 K \\ 
  \hline
 1  & 1 & -1067.6 & 36.56 & 1 & -1134.4 & 37.00 \\
 2  & 3 &  -845.2 & 17.22 & 3 &  -914.7 & 17.77 \\
 3  & 2 &  -785.9 &  8.03 & 2 &  -850.2 &  7.98 \\
 4  & 1 &  -771.3 &  4.34 & 1 &  -829.9 &  4.14 \\
 5  & 3 &  -697.5 &  5.95 & 3 &  -742.5 &  5.15 \\
 6  & 2 &  -616.0 &  2.37 & 2 &  -687.8 &  2.48 \\
 7  & 5 &  -613.1 &  5.10 & 5 &  -685.5 &  5.37 \\
 8  & 1 &  -590.0 &  1.18 & 1 &  -667.0 &  1.28 \\
 9  & 3 &  -580.3 &  2.56 & 3 &  -650.1 &  2.65 \\
 10 & 2 &  -535.1 &  1.32 & 2 &  -595.0 &  1.27 \\
 11 & 4 &  -531.4 &  2.06 & 4 &  -594.0 &  2.02 \\
 12 & 1 &  -525.4 &  0.74 & 1 &  -590.6 &  0.74 \\
 13 & 3 &  -485.2 &  1.29 & 3 &  -543.5 &  1.23 \\
 14 & 5 &  -464.8 &  1.75 & 5 &  -515.2 &  1.58 \\
 15 & 4 &  -445.7 &  1.11 & 4 &  -495.6 &  1.00 \\
 16 & 2 &  -404.6 &  0.52 & 2 &  -456.7 &  0.47 \\
 17 & 7 &  -371.4 &  1.22 & 7 &  -446.6 &  1.31 \\
 18 & 4 &  -362.7 &  0.61 & 3 &  -436.7 &  0.57 \\
 19 & 3 &  -362.1 &  0.53 & 4 &  -432.7 &  0.63 \\
 20 & 5 &  -339.6 &  0.71 & 5 &  -414.8 &  0.77 \\
 21 & 0 &  -330.5 &  0.06 & 0 &  -400.2 &  0.06 \\
 22 & 2 &  -317.4 &  0.28 & 4 &  -388.3 &  0.46 \\
 23 & 4 &  -316.8 &  0.44 & 2 &  -387.9 &  0.29 \\
 24 & 1 &  -291.3 &  0.14 & 1 &  -357.6 &  0.14 \\
 25 & 6 &  -275.4 &  0.53 & 6 &  -336.5 &  0.52 \\
  \hline
  Sum & & & 96.62 & & & 96.88  \\ 
 \end{tabular}
\end{ruledtabular}
\end{table}
Simulating the spectrum of Gunnarsson {\it et al}., we encounter two problems.
First, the spectrum is too broad and, second, the vibrational temperature 
is too high.
In fact, the statistical weights of the ground vibronic level at 200 K
are about 37 \% in both cases (see Table \ref{TABLE:VIBRO_WEIGHT_GUNNARSSON}), 
hence, we must consider many excited vibronic states. 
To represent the excited vibronic states with enough accuracy, 
we expect that the vibronic basis must be larger than the present one.
Furthermore, as the vibrational temperature is further increased, 
the weight of each vibronic level and the shape of the spectrum varies 
easily.
Although the range of the vibrational temperature is not reported, 
we increased the temperature by 20 K which is the uncertainty range 
of vibrational temperature in the case of Wang {\it et al}. \cite{Wang2005a}
The statistical weights of the ground vibronic level decreased from 
ca. 37 \% to ca. 31 \% with this increase of the temperature.
This change of the weight affects the shape of the spectrum.
Therefore it is difficult to perform an accurate simulation and to 
estimate VCCs from the spectrum of Gunnarsson {\it et al}. 

Given better the experimental conditions of Wang {\it et al}. \cite{Wang2005a}
allowing for more accurate simulations, 
we may conclude that the VCCs extracted from these experiments
should be considered more reliable than those obtained by 
Gunnarsson {\it et al}. \cite{Gunnarsson1995a}

\subsection{DFT calculations of the vibronic coupling constants
\label{SUBSEC:CALC_VCC}}
We compute the vibronic coupling constants of C$_{60}^-$ 
using the DFT method described in Sec. \ref{SEC:DFTVCC}.
{
For DFT calculations with pure functionals,
it is well known that in the high-symmetry geometry the occupied level 
belonging to one degenerate representation
moves upwards in energy relative to the empty levels
belonging to the same degenerate manifold. \cite{Bruyndonckx1997a}
However in our case the situation is opposite (the occupied level
is the lowest one) because of the Hartree--Fock exchange contribution
contained in the B3LYP functional.
The splitting between the $t_{1u}$ Kohn--Sham levels are about 1 eV 
(see Table \ref{TABLE:KSGAP}) and the variations of the total 
electronic energies for different occupation schemes of $t_{1u}$ orbitals
are less than 0.2 meV.
Moreover, the vibronic coupling constants do not depend on the choice of the
electronic states significantly. 
The variation of the total stabilization energy is ca. 1 meV.
}
\begin{table}
{
\caption{The splitting between the lifted one-electron $t_{1u}$ levels 
$\Delta \epsilon$ (meV).
\label{TABLE:KSGAP}}
}
\begin{ruledtabular}
{
 \begin{tabular}{ccccccc}
  & \multicolumn{3}{c}{6-311G(d)} & 6-311+G(d) & \multicolumn{2}{c}{cc-pVTZ} \\
  & 20 \% & 25 \% & 30 \% & 20 \% & 20 \% & 25 \% \\
  \hline
  $\Delta \epsilon$ & 691 & 874 & 1059 & 686 & 693 & 876 \\
 \end{tabular}
}
\end{ruledtabular}
\end{table}
The dimensionless VCCs and the stabilization energies 
are shown in Table \ref{TABLE:VCC} and \ref{TABLE:PRESENT_STABILIZATION}.
Although we use several basis sets, the VCCs do not depend on the basis set 
significantly.
On the other hand, the VCCs vary with the increase of the fraction of 
the Hartree--Fock exchange energy in the exchange-correlation functional.
Increasing this fraction leads to larger VCCs and stabilization energies.
We find that the stabilization energies of high frequency modes,
$a_g(2), h_g(7)$, and $h_g(8)$ are the strongest. 
Compared with other DFT calculations, we may conclude that 
the stabilization energies of $a_g$ modes agree well with 
the previous calculations,
while the present stabilization energy $E_s + E_{\rm JT}$ 
is larger than the previous results.

In comparison with present simulation of the experimental PES, 
the DFT calculations with the energy functionals including fractions 
of 20 \% and 25 \% of the Hartree--Fock exchange energy 
give close values.
Although the stabilization energy $E_s + E_{\rm JT}$ 
obtained using the original B3LYP functional is slightly 
smaller than the experimental value, 
the result obtained with it 
is also close to the experimental one.
The distribution of the computed stabilization energy of each $h_g$ mode 
$E_{{\rm JT},\mu}$ qualitatively agrees with the experimental 
results.
The stabilization energy of the $h_g(7)$ mode is obtained 
smaller and that of the 
$h_g(8)$ mode is obtained larger than the experimental values.
The slight difference between theoretical and experimental results 
should originate from still inaccurately computed vibrational vectors.
Indeed, it was shown that a small mixing of the vibrational vectors in 
fullerene affects the values of VCCs significantly.
\cite{Gunnarsson1995a}
Besides the present computational results, 
the LDA calculation by Manini {\it et al}.\cite{Manini2001a} and 
the GGA calculation by Frederiksen {\it et al}.\cite{Frederiksen2008a}
give similar relative values for the coupling 
constants of $h_g$ modes 
as the present simulations of PES of Wang {\it et al}. \cite{Wang2005a}
On the other hand, these calculations give smaller absolute values of 
the VCCs and of the total stabilization energies 
than the presently obtained. 
{
Contrary to these calculations, the B3LYP calculation by Saito 
\cite{Saito2002a} and Laflamme Janssen \cite{Janssen2010a} 
give close values of VCCs to the present results. 
However, despite of using the same B3LYP functional, our calculations
of VCCs differ from the ones in Refs. \onlinecite{Saito2002a} and 
\onlinecite{Janssen2010a} since we used here the derivative of the total 
energy, which does not coincide with the derivative of the Kohn--Sham
orbital energy (see the appendix).
}
The LDA calculation of Breda {\it et al}.\cite{Breda1998a} 
gives the distribution of relative strengths of VCCs
which is similar to the results of Gunnarsson {\it et al}. 
and do not agree with the values derived here from PES of Wang {\it et al}. 
Varma {\it et al}.\cite{Varma1991a} computed VCCs using MNDO method, 
however, the distribution of the stabilization energies is different from 
the present simulations of experiment and the theoretical values obtained here. 

On the contrary, the theoretical stabilization energy of 
the $a_g(1)$ mode is too small and that of the $a_g(2)$ mode is too large
compared to $E_{s,i}$ derived from experiment (Tables \ref{TABLE:VCC}, 
\ref{TABLE:PRESENT_STABILIZATION}).
To find the correct order of the corresponding VCCs, we derive the coupling 
constants of the $a_g$ modes from the experimental bond lengths
as described in Sect. \ref{SEC:C-C}.
The obtained stabilization energies $E_{s,i}$ 
are shown in Table \ref{TABLE:VCC_BOND_LENGTH}.
\begin{table}
\caption{Stabilization energies (meV) derived from the experimental 
C-C bond lengths of C$_{60}$ and C$_{60}^-$.
\label{TABLE:VCC_BOND_LENGTH}}
 \begin{ruledtabular}
 \begin{tabular}{ccccccc}
   & Frequency & \multicolumn{5}{c}{Stabilization energies of $a_g$ modes}\\
   &(cm$^{-1}$)& \multicolumn{5}{c}{(meV)} \\
   &           & (3) & (10)
   & Narymbetov\footnotemark[1] 
   & Fujiwara\footnotemark[1] & Fujiwara\footnotemark[2]\\
   &  &  &  & 7K & 25K & 90K \\
  \hline
   $a_g(1)$ & 496 & 7.7 & 2.6 & 8.0 & 0.1 & 58.8\\
   $a_g(2)$ &1470 & 8.2 &16.9 & 0.0 & 7.5 & 7.0\\
 \end{tabular}
 \end{ruledtabular}
\footnotetext[1]{The structure of C$_{60}$ is taken from neutron diffraction 
at 5K (Ref.~\onlinecite{David1991a}).}
\footnotetext[2]{The structure of C$_{60}$ is taken from X-ray diffraction 
at 110K (Ref.~\onlinecite{Burgi1992a}).}
\end{table}
Unfortunately, as we can see, $E_{s,i}$ depend very strongly on the set 
of the C-C bond lengths, thus we cannot draw a conclusion about their 
relative strength.
This comes from the fact that we use the structural data measured by different 
techniques (X-ray and neutron scattering) on different systems 
(TDAE-C$_{60}$ and fullerite) for C$_{60}^-$ and C$_{60}$, respectively.
In both cases, fullerenes should be deformed due to the environment 
compared with the free C$_{60}$ molecule. 
The distortions caused by the crystal fields of C$_{60}$ in 
TDAE-C$_{60}$ and fullerite are different from each other.
Furthermore, 
the expected changes in the bond lengths in C$_{60}$ and C$_{60}^-$ 
are within the experimental accuracy of the structural data.
Thus, the origin of the discrepancy of the relative strength of VCCs 
of the $a_g$ modes between 
the simulations and DFT calculations remains unclear.

Note that
although the values of $E_{s,i}$ are different from the experimental 
results, their sum, as well as, the distribution of $E_{{\rm JT},\mu}$ 
and $E_s+E_{\rm JT}$ 
are close to the experimental values.
Therefore, we may conclude that the present theoretical method gives 
improved values of vibronic coupling constants.

\section{CONCLUSION
\label{SEC:CONCLUSION}}
In this work, we simulated the PES of Wang {\it et al}.
and derived the vibronic coupling constants of C$_{60}^-$. 
We obtain several sets of VCCs, because the frequencies of $a_g(2)$, $h_g(7)$, 
and $h_g(8)$ modes are close to each other.
Considering the ambiguity of the vibrational temperature in the 
experiment, these sets of VCCs cannot be distinguished.
Thus, to obtain more accurate coupling constants, it is desired to 
perform an observation of a PES of C$_{60}^-$ in still better
experimental conditions.
Although we find several sets of VCCs from the spectrum,
the stabilization energies are similar to each other. 
In comparison with the total stabilization energy derived by 
Gunnarsson {\it et al}., \cite{Gunnarsson1995a} our value is smaller by 30 \%.
We also calculated the VCCs using the DFT method.
Even though the experimental and theoretical orders of $E_{s,i}$ 
disagree with each other, the distribution of $E_{{\rm JT},\mu}$ and 
the total stabilization energy $E_s + E_{\rm JT}$ agrees well 
with the experimental values.
Thus we may conclude that the problem of the discrepancy between the 
experimental and calculated coupling constants, persistent in the 
previous studies, is basically solved in the present work.
As an extension of the present work we expect that the theoretical 
approach used here
could be successfully applied for the calculation of VCC of C$_{60}^{n-}$ 
anions in A$_n$C$_{60}$ fullerides as well as of their multiplet 
splitting parameters.

\appendix*
{
\section{Vibronic coupling constants and the gradient of Kohn--Sham levels}
The total energy $E(\mathbf{R})$ in the DFT is written as follows:
\begin{eqnarray}
 E(\mathbf{R}) &=& \sum_{\mu \Gamma \gamma}
 \epsilon_{\mu \Gamma \gamma}(\mathbf{R}) 
 - \frac{1}{2}\int d\mathbf{r} 
    \int d\mathbf{r}' 
 \frac{n(\mathbf{r}; \mathbf{R}) n(\mathbf{r}'; \mathbf{R})}
{\left|\mathbf{r} - \mathbf{r}' \right|}
\notag \\
 &+& E_{xc}\left[n(\mathbf{r}; \mathbf{R})\right]
 - \int d\mathbf{r} n(\mathbf{r}; \mathbf{R}) V_{xc}(\mathbf{r}; \mathbf{R})
 + V_{nn}(\mathbf{R}).
\nonumber\\
\label{Eq:TotE}
\end{eqnarray}
Here, 
$\Gamma \gamma$ is the irreducible representation of the Kohn--Sham 
orbital, $\mu$ is the quantum number other than $\Gamma \gamma$, 
$\epsilon_{\mu \Gamma \gamma}$ is the Kohn--Sham level, 
$\sum_{\mu \Gamma \gamma}$ is taken over occupied levels, 
$n(\mathbf{r}; \mathbf{R})$ is the ground electronic density,
$E_{xc}\left[n(\mathbf{r}; \mathbf{R})\right]$ is 
the exchange-correlation energy functional, 
$V_{xc}(\mathbf{r}, \mathbf{R})$ is the exchange-correlation potential,
and $V_{nn}(\mathbf{R})$ is the Coulomb potential energy between nuclei. 
The vibronic coupling constant of $\Gamma' \gamma'$ mode is 
\begin{widetext}
\begin{eqnarray}
 V_{\Gamma' \gamma'} 
 &=& 
 \sum_{\mu \Gamma \gamma}
 \left(\frac{\partial \epsilon_{\mu \Gamma \gamma}(\mathbf{R})}
 {\partial Q_{\Gamma' \gamma'}} \right)_{\mathbf{R}_0}
 - \int d\mathbf{r} 
 \left(\frac{\partial n(\mathbf{r}; \mathbf{R})}
 {\partial Q_{\Gamma' \gamma'}}\right)_{\mathbf{R}_0}
 \int d\mathbf{r}' \frac{n(\mathbf{r}'; \mathbf{R})}
{\left|\mathbf{r} - \mathbf{r}' \right|}
\notag \\
 &-& 
 \int d\mathbf{r} n(\mathbf{r}; \mathbf{R}) 
 \left(\frac{\partial V_{xc}(\mathbf{r}; \mathbf{R})}
 {\partial Q_{\Gamma' \gamma'}}\right)_{\mathbf{R}_0} + 
 \left(\frac{\partial V_{nn}(\mathbf{R})}{\partial Q_{\Gamma' \gamma'}}
 \right)_{\mathbf{R}_0}.
\label{Eq:der_TotE}
\end{eqnarray}
\end{widetext}
For the totally symmetric modes, all the derivatives in the right-hand side
of Eq. (\ref{Eq:der_TotE}) are not zero.
For the Jahn--Teller active modes, the sum of the gradient of the completely
occupied Kohn--Sham levels belonging to the same $\Gamma$ is zero due to 
the symmetry reasons, and the gradient of the 
Coulomb potential between the nuclei is zero also because of the symmetry.
However, the second and third terms which include the derivative of 
$n(\mathbf{r}; \mathbf{R})$ and $V_{xc}(\mathbf{r}; \mathbf{R})$ with 
respect to $Q_{\Gamma' \gamma'}$, respectively, are nonzero
because $n(\mathbf{r}; \mathbf{R})$ is not a totally symmetric function.
Therefore, in general, the vibronic coupling constant is not equal to 
the gradient of the frontier Kohn--Sham level.
}

\begin{acknowledgments}
We would like to thank Dr. X. B. Wang and Prof. L. S. Wang for sending us 
unpublished data.
N.I. would like to thank the research fund for study abroad
from the Research Project of Nano Frontier, 
graduate school of engineering, Kyoto University.
T.S. and N.I. are grateful to the Division of Quantum and Physical 
Chemistry at the University of Leuven for hospitality.
Theoretical calculations were partly performed using Research Center 
for Computational Science, Okazaki, Japan.
This work was supported by a
Grant-in-Aid for Scientific Research, priority area
‘Molecular theory for real systems’ (20038028) from the Japan
Society for the Promotion of Science (JSPS).
This work was also supported
in part by the Global COE Program ‘International Center for
Integrated Research and Advanced Education in Materials Science’
(No. B-09) of the Ministry of Education, Culture, Sports, Science and
Technology (MEXT) of Japan, administered by the JSPS.
Financial support from the JSPS--FWO (Fonds voor Wetenschappelijk 
Onderzoek-Vlaanderen) bilateral program is gratefully 
acknowledged. 
\end{acknowledgments}


\begin{thebibliography}{34}%
\makeatletter
\providecommand \@ifxundefined [1]{%
 \@ifx{#1\undefined}
}%
\providecommand \@ifnum [1]{%
 \ifnum #1\expandafter \@firstoftwo
 \else \expandafter \@secondoftwo
 \fi
}%
\providecommand \@ifx [1]{%
 \ifx #1\expandafter \@firstoftwo
 \else \expandafter \@secondoftwo
 \fi
}%
\providecommand \natexlab [1]{#1}%
\providecommand \enquote  [1]{``#1''}%
\providecommand \bibnamefont  [1]{#1}%
\providecommand \bibfnamefont [1]{#1}%
\providecommand \citenamefont [1]{#1}%
\providecommand \href@noop [0]{\@secondoftwo}%
\providecommand \href [0]{\begingroup \@sanitize@url \@href}%
\providecommand \@href[1]{\@@startlink{#1}\@@href}%
\providecommand \@@href[1]{\endgroup#1\@@endlink}%
\providecommand \@sanitize@url [0]{\catcode `\\12\catcode `\$12\catcode
  `\&12\catcode `\#12\catcode `\^12\catcode `\_12\catcode `\%12\relax}%
\providecommand \@@startlink[1]{}%
\providecommand \@@endlink[0]{}%
\providecommand \url  [0]{\begingroup\@sanitize@url \@url }%
\providecommand \@url [1]{\endgroup\@href {#1}{\urlprefix }}%
\providecommand \urlprefix  [0]{URL }%
\providecommand \Eprint [0]{\href }%
\@ifxundefined \urlstyle {%
  \providecommand \doi  [0]{\begingroup \@sanitize@url \@doi}%
  \providecommand \@doi [1]{\endgroup \@@startlink {\doibase
  #1}doi:\discretionary {}{}{}#1\@@endlink }%
}{%
  \providecommand \doi  [0]{doi:\discretionary{}{}{}\begingroup
  \urlstyle{rm}\Url }%
}%
\providecommand \doibase [0]{http://dx.doi.org/}%
\providecommand \Doi [0]{\begingroup \@sanitize@url \@Doi }%
\providecommand \@Doi  [1]{\endgroup\@@startlink{\doibase#1}\@@Doi}%
\providecommand \@@Doi [1]{#1\@@endlink}%
\providecommand \selectlanguage [0]{\@gobble}%
\providecommand \bibinfo  [0]{\@secondoftwo}%
\providecommand \bibfield  [0]{\@secondoftwo}%
\providecommand \translation [1]{[#1]}%
\providecommand \BibitemOpen [0]{}%
\providecommand \bibitemStop [0]{}%
\providecommand \bibitemNoStop [0]{.\EOS\space}%
\providecommand \EOS [0]{\spacefactor3000\relax}%
\providecommand \BibitemShut  [1]{\csname bibitem#1\endcsname}%
\bibitem [{\citenamefont {Chancey}\ and\ \citenamefont
  {O'Brien}(1997)}]{Chancey1997a}%
  \BibitemOpen
  \bibfield  {author} {\bibinfo {author} {\bibfnamefont {C.~C.}\ \bibnamefont
  {Chancey}}\ and\ \bibinfo {author} {\bibfnamefont {M.~C.~M.}\ \bibnamefont
  {O'Brien}},\ }\href@noop {} {\emph {\bibinfo {title} {The Jahn--Teller Effect
  in C$_{60}$ and other Icosahedral Complexes}}}\ (\bibinfo  {publisher}
  {Princeton University Press},\ \bibinfo {address} {Princeton},\ \bibinfo
  {year} {1997})\BibitemShut {NoStop}%
\bibitem [{\citenamefont {Gunnarsson}(2004)}]{Gunnarsson2004a}%
  \BibitemOpen
  \bibfield  {author} {\bibinfo {author} {\bibfnamefont {O.}~\bibnamefont
  {Gunnarsson}},\ }\href@noop {} {\emph {\bibinfo {title} {Alkali-Doped
  Fullerides: Narrow-Band Solids with Unusual Properties}}}\ (\bibinfo
  {publisher} {World Scientific},\ \bibinfo {address} {Singapore},\ \bibinfo
  {year} {2004})\BibitemShut {NoStop}%
\bibitem [{\citenamefont {Gunnarsson}\ \emph {et~al.}(1995)\citenamefont
  {Gunnarsson}, \citenamefont {Handschuh}, \citenamefont {Bechthold},
  \citenamefont {Kessler}, \citenamefont {Gantef{\"{o}}r},\ and\ \citenamefont
  {Eberhardt}}]{Gunnarsson1995a}%
  \BibitemOpen
  \bibfield  {author} {\bibinfo {author} {\bibfnamefont {O.}~\bibnamefont
  {Gunnarsson}}, \bibinfo {author} {\bibfnamefont {H.}~\bibnamefont
  {Handschuh}}, \bibinfo {author} {\bibfnamefont {P.~S.}\ \bibnamefont
  {Bechthold}}, \bibinfo {author} {\bibfnamefont {B.}~\bibnamefont {Kessler}},
  \bibinfo {author} {\bibfnamefont {G.}~\bibnamefont {Gantef{\"{o}}r}}, \ and\
  \bibinfo {author} {\bibfnamefont {W.}~\bibnamefont {Eberhardt}},\ }\href@noop
  {} {\bibfield  {journal} {\bibinfo  {journal} {Phys. Rev. Lett.},\ }\textbf
  {\bibinfo {volume} {74}},\ \bibinfo {pages} {1875} (\bibinfo {year}
  {1995})}\BibitemShut {NoStop}%
\bibitem [{\citenamefont {Winter}\ and\ \citenamefont
  {Kuzmany}(1996)}]{Winter1996a}%
  \BibitemOpen
  \bibfield  {author} {\bibinfo {author} {\bibfnamefont {J.}~\bibnamefont
  {Winter}}\ and\ \bibinfo {author} {\bibfnamefont {H.}~\bibnamefont
  {Kuzmany}},\ }\href@noop {} {\bibfield  {journal} {\bibinfo  {journal} {Phys.
  Rev. B},\ }\textbf {\bibinfo {volume} {53}},\ \bibinfo {pages} {655}
  (\bibinfo {year} {1996})}\BibitemShut {NoStop}%
{
\bibitem [{\citenamefont {Hands}\ \emph {et~al.}(2008)\citenamefont {Hands},
  \citenamefont {Dunn}, \citenamefont {Bates}, \citenamefont {Hope},
  \citenamefont {Meech},\ and\ \citenamefont {Andrews}}]{Hands2008a}%
  \BibitemOpen
  \bibfield  {author} {\bibinfo {author} {\bibfnamefont {I.~D.}\ \bibnamefont
  {Hands}}, \bibinfo {author} {\bibfnamefont {J.~L.}\ \bibnamefont {Dunn}},
  \bibinfo {author} {\bibfnamefont {C.~A.}\ \bibnamefont {Bates}}, \bibinfo
  {author} {\bibfnamefont {M.~J.}\ \bibnamefont {Hope}}, \bibinfo {author}
  {\bibfnamefont {S.~R.}\ \bibnamefont {Meech}}, \ and\ \bibinfo {author}
  {\bibfnamefont {D.~L.}\ \bibnamefont {Andrews}},\ }\href@noop {} {\bibfield
  {journal} {\bibinfo  {journal} {Phys. Rev. B},\ }\textbf {\bibinfo {volume}
  {77}},\ \bibinfo {pages} {115445} (\bibinfo {year} {2008})}\BibitemShut
  {NoStop}%
}
\bibitem [{\citenamefont {Varma}\ \emph {et~al.}(1991)\citenamefont {Varma},
  \citenamefont {Zaanen},\ and\ \citenamefont {Raghavachari}}]{Varma1991a}%
  \BibitemOpen
  \bibfield  {author} {\bibinfo {author} {\bibfnamefont {C.~M.}\ \bibnamefont
  {Varma}}, \bibinfo {author} {\bibfnamefont {J.}~\bibnamefont {Zaanen}}, \
  and\ \bibinfo {author} {\bibfnamefont {K.}~\bibnamefont {Raghavachari}},\
  }\href@noop {} {\bibfield  {journal} {\bibinfo  {journal} {Science},\
  }\textbf {\bibinfo {volume} {254}},\ \bibinfo {pages} {989} (\bibinfo {year}
  {1991})}\BibitemShut {NoStop}%
\bibitem [{\citenamefont {Schluter}\ \emph {et~al.}(1992)\citenamefont
  {Schluter}, \citenamefont {Lannoo}, \citenamefont {Needels}, \citenamefont
  {Baraff},\ and\ \citenamefont {Tom{\'{a}}nek}}]{Schluter1992a}%
  \BibitemOpen
  \bibfield  {author} {\bibinfo {author} {\bibfnamefont {M.}~\bibnamefont
  {Schluter}}, \bibinfo {author} {\bibfnamefont {M.}~\bibnamefont {Lannoo}},
  \bibinfo {author} {\bibfnamefont {M.}~\bibnamefont {Needels}}, \bibinfo
  {author} {\bibfnamefont {G.~A.}\ \bibnamefont {Baraff}}, \ and\ \bibinfo
  {author} {\bibfnamefont {D.}~\bibnamefont {Tom{\'{a}}nek}},\ }\href@noop {}
  {\bibfield  {journal} {\bibinfo  {journal} {Phys. Rev. Lett.},\ }\textbf
  {\bibinfo {volume} {68}},\ \bibinfo {pages} {526} (\bibinfo {year}
  {1992})}\BibitemShut {NoStop}%
\bibitem [{\citenamefont {Faulhaber}\ \emph {et~al.}(1993)\citenamefont
  {Faulhaber}, \citenamefont {Ko},\ and\ \citenamefont
  {Briddon}}]{Faulhaber1993a}%
  \BibitemOpen
  \bibfield  {author} {\bibinfo {author} {\bibfnamefont {J.~C.~R.}\
  \bibnamefont {Faulhaber}}, \bibinfo {author} {\bibfnamefont {D.~Y.~K.}\
  \bibnamefont {Ko}}, \ and\ \bibinfo {author} {\bibfnamefont {P.~R.}\
  \bibnamefont {Briddon}},\ }\href@noop {} {\bibfield  {journal} {\bibinfo
  {journal} {Phys. Rev. B},\ }\textbf {\bibinfo {volume} {48}},\ \bibinfo
  {pages} {661} (\bibinfo {year} {1993})}\BibitemShut {NoStop}%
\bibitem [{\citenamefont {Antropov}\ \emph {et~al.}(1993)\citenamefont
  {Antropov}, \citenamefont {Gunnarsson},\ and\ \citenamefont
  {Liechtenstein}}]{Antropov1993a}%
  \BibitemOpen
  \bibfield  {author} {\bibinfo {author} {\bibfnamefont {V.~P.}\ \bibnamefont
  {Antropov}}, \bibinfo {author} {\bibfnamefont {O.}~\bibnamefont
  {Gunnarsson}}, \ and\ \bibinfo {author} {\bibfnamefont {A.~I.}\ \bibnamefont
  {Liechtenstein}},\ }\href@noop {} {\bibfield  {journal} {\bibinfo  {journal}
  {Phys. Rev. B},\ }\textbf {\bibinfo {volume} {48}},\ \bibinfo {pages} {7651}
  (\bibinfo {year} {1993})}\BibitemShut {NoStop}%
\bibitem [{\citenamefont {Breda}\ \emph {et~al.}(1998)\citenamefont {Breda},
  \citenamefont {Broglia}, \citenamefont {Col{\`{o}}}, \citenamefont {Roman},
  \citenamefont {Alasia}, \citenamefont {Onida}, \citenamefont {Ponomarev},\
  and\ \citenamefont {Vigezzi}}]{Breda1998a}%
  \BibitemOpen
  \bibfield  {author} {\bibinfo {author} {\bibfnamefont {N.}~\bibnamefont
  {Breda}}, \bibinfo {author} {\bibfnamefont {R.~A.}\ \bibnamefont {Broglia}},
  \bibinfo {author} {\bibfnamefont {G.}~\bibnamefont {Col{\`{o}}}}, \bibinfo
  {author} {\bibfnamefont {H.~E.}\ \bibnamefont {Roman}}, \bibinfo {author}
  {\bibfnamefont {F.}~\bibnamefont {Alasia}}, \bibinfo {author} {\bibfnamefont
  {G.}~\bibnamefont {Onida}}, \bibinfo {author} {\bibfnamefont
  {V.}~\bibnamefont {Ponomarev}}, \ and\ \bibinfo {author} {\bibfnamefont
  {E.}~\bibnamefont {Vigezzi}},\ }\href@noop {} {\bibfield  {journal} {\bibinfo
   {journal} {Chem. Phys. Lett.},\ }\textbf {\bibinfo {volume} {286}},\
  \bibinfo {pages} {350} (\bibinfo {year} {1998})}\BibitemShut {NoStop}%
\bibitem [{\citenamefont {Manini}\ \emph {et~al.}(2001)\citenamefont {Manini},
  \citenamefont {Carso}, \citenamefont {Fabrizio},\ and\ \citenamefont
  {Tosatti}}]{Manini2001a}%
  \BibitemOpen
  \bibfield  {author} {\bibinfo {author} {\bibfnamefont {N.}~\bibnamefont
  {Manini}}, \bibinfo {author} {\bibfnamefont {A.~D.}\ \bibnamefont {Carso}},
  \bibinfo {author} {\bibfnamefont {M.}~\bibnamefont {Fabrizio}}, \ and\
  \bibinfo {author} {\bibfnamefont {E.}~\bibnamefont {Tosatti}},\ }\href@noop
  {} {\bibfield  {journal} {\bibinfo  {journal} {Philos. Mag. B},\ }\textbf
  {\bibinfo {volume} {81}},\ \bibinfo {pages} {793} (\bibinfo {year}
  {2001})}\BibitemShut {NoStop}%
\bibitem [{\citenamefont {Saito}(2002)}]{Saito2002a}%
  \BibitemOpen
  \bibfield  {author} {\bibinfo {author} {\bibfnamefont {M.}~\bibnamefont
  {Saito}},\ }\href@noop {} {\bibfield  {journal} {\bibinfo  {journal} {Phys.
  Rev. B},\ }\textbf {\bibinfo {volume} {65}},\ \bibinfo {pages} {220508(R)}
  (\bibinfo {year} {2002})}\BibitemShut {NoStop}%
\bibitem [{\citenamefont {Frederiksen}\ \emph {et~al.}(2008)\citenamefont
  {Frederiksen}, \citenamefont {Franke}, \citenamefont {Arnau}, \citenamefont
  {Schulze}, \citenamefont {Pascual},\ and\ \citenamefont
  {Lorente}}]{Frederiksen2008a}%
  \BibitemOpen
  \bibfield  {author} {\bibinfo {author} {\bibfnamefont {T.}~\bibnamefont
  {Frederiksen}}, \bibinfo {author} {\bibfnamefont {K.~J.}\ \bibnamefont
  {Franke}}, \bibinfo {author} {\bibfnamefont {A.}~\bibnamefont {Arnau}},
  \bibinfo {author} {\bibfnamefont {G.}~\bibnamefont {Schulze}}, \bibinfo
  {author} {\bibfnamefont {J.~I.}\ \bibnamefont {Pascual}}, \ and\ \bibinfo
  {author} {\bibfnamefont {N.}~\bibnamefont {Lorente}},\ }\href@noop {}
  {\bibfield  {journal} {\bibinfo  {journal} {Phys. Rev. B},\ }\textbf
  {\bibinfo {volume} {78}},\ \bibinfo {pages} {233401} (\bibinfo {year}
  {2008})}\BibitemShut {NoStop}%
\bibitem [{\citenamefont {Laflamme~Janssen}\ \emph {et~al.}(2010)\citenamefont
  {Laflamme~Janssen}, \citenamefont {C\^ot\'e}, \citenamefont {Louie},\ and\
  \citenamefont {Cohen}}]{Janssen2010a}%
  \BibitemOpen
  \bibfield  {author} {\bibinfo {author} {\bibfnamefont {J.}~\bibnamefont
  {Laflamme~Janssen}}, \bibinfo {author} {\bibfnamefont {M.}~\bibnamefont
  {C\^ot\'e}}, \bibinfo {author} {\bibfnamefont {S.~G.}\ \bibnamefont {Louie}},
  \ and\ \bibinfo {author} {\bibfnamefont {M.~L.}\ \bibnamefont {Cohen}},\
  }\Doi {10.1103/PhysRevB.81.073106} {\bibfield  {journal} {\bibinfo  {journal}
  {Phys. Rev. B},\ }\textbf {\bibinfo {volume} {81}},\ \bibinfo {pages}
  {073106} (\bibinfo {year} {2010})}\BibitemShut {NoStop}%
\bibitem [{\citenamefont {Wang}\ \emph {et~al.}(2005)\citenamefont {Wang},
  \citenamefont {Woo},\ and\ \citenamefont {Wang}}]{Wang2005a}%
  \BibitemOpen
  \bibfield  {author} {\bibinfo {author} {\bibfnamefont {X.~B.}\ \bibnamefont
  {Wang}}, \bibinfo {author} {\bibfnamefont {H.~K.}\ \bibnamefont {Woo}}, \
  and\ \bibinfo {author} {\bibfnamefont {L.~S.}\ \bibnamefont {Wang}},\
  }\href@noop {} {\bibfield  {journal} {\bibinfo  {journal} {J. Chem. Phys.},\
  }\textbf {\bibinfo {volume} {123}},\ \bibinfo {pages} {051106} (\bibinfo
  {year} {2005})}\BibitemShut {NoStop}%
\bibitem [{\citenamefont {O'Brien}(1971)}]{OBrien1971a}%
  \BibitemOpen
  \bibfield  {author} {\bibinfo {author} {\bibfnamefont {M.~C.~M.}\
  \bibnamefont {O'Brien}},\ }\href@noop {} {\bibfield  {journal} {\bibinfo
  {journal} {J. Phys. C: Solid St. Phys.},\ }\textbf {\bibinfo {volume} {4}},\
  \bibinfo {pages} {2524} (\bibinfo {year} {1971})}\BibitemShut {NoStop}%
\bibitem [{\citenamefont {Auerbach}\ \emph {et~al.}(1994)\citenamefont
  {Auerbach}, \citenamefont {Manini},\ and\ \citenamefont
  {Tosatti}}]{Auerbach1994a}%
  \BibitemOpen
  \bibfield  {author} {\bibinfo {author} {\bibfnamefont {A.}~\bibnamefont
  {Auerbach}}, \bibinfo {author} {\bibfnamefont {N.}~\bibnamefont {Manini}}, \
  and\ \bibinfo {author} {\bibfnamefont {E.}~\bibnamefont {Tosatti}},\
  }\href@noop {} {\bibfield  {journal} {\bibinfo  {journal} {Phys. Rev. B},\
  }\textbf {\bibinfo {volume} {49}},\ \bibinfo {pages} {12998} (\bibinfo {year}
  {1994})}\BibitemShut {NoStop}%
\bibitem [{\citenamefont {Edmonds}(1974)}]{Edmonds1974a}%
  \BibitemOpen
  \bibfield  {author} {\bibinfo {author} {\bibfnamefont {A.~R.}\ \bibnamefont
  {Edmonds}},\ }\href@noop {} {\emph {\bibinfo {title} {Angular Momentum in
  Quantum Mechanics}}}\ (\bibinfo  {publisher} {Princeton University Press},\
  \bibinfo {address} {Princeton},\ \bibinfo {year} {1974})\BibitemShut
  {NoStop}%
\bibitem [{\citenamefont {O'Brien}(1969)}]{OBrien1969a}%
  \BibitemOpen
  \bibfield  {author} {\bibinfo {author} {\bibfnamefont {M.~C.~M.}\
  \bibnamefont {O'Brien}},\ }\href@noop {} {\bibfield  {journal} {\bibinfo
  {journal} {Phys. Rev.},\ }\textbf {\bibinfo {volume} {187}},\ \bibinfo
  {pages} {407} (\bibinfo {year} {1969})}\BibitemShut {NoStop}%
\bibitem [{\citenamefont {Bersuker}\ and\ \citenamefont
  {Polinger}(1989)}]{Bersuker1989a}%
  \BibitemOpen
  \bibfield  {author} {\bibinfo {author} {\bibfnamefont {I.~B.}\ \bibnamefont
  {Bersuker}}\ and\ \bibinfo {author} {\bibfnamefont {V.~Z.}\ \bibnamefont
  {Polinger}},\ }\href@noop {} {\emph {\bibinfo {title} {Vibronic Interactions
  in Molecules and Crystals}}}\ (\bibinfo  {publisher} {Springer--Verlag},\
  \bibinfo {address} {Berlin and Heidelberg},\ \bibinfo {year}
  {1989})\BibitemShut {NoStop}%
\bibitem [{\citenamefont {Bethune}\ \emph {et~al.}(1991)\citenamefont
  {Bethune}, \citenamefont {Meijer}, \citenamefont {Tang}, \citenamefont
  {Rosen}, \citenamefont {Golden}, \citenamefont {Seki}, \citenamefont
  {Brown},\ and\ \citenamefont {de~Vries}}]{Bethune1991a}%
  \BibitemOpen
  \bibfield  {author} {\bibinfo {author} {\bibfnamefont {D.~S.}\ \bibnamefont
  {Bethune}}, \bibinfo {author} {\bibfnamefont {G.}~\bibnamefont {Meijer}},
  \bibinfo {author} {\bibfnamefont {W.~C.}\ \bibnamefont {Tang}}, \bibinfo
  {author} {\bibfnamefont {H.~J.}\ \bibnamefont {Rosen}}, \bibinfo {author}
  {\bibfnamefont {W.~G.}\ \bibnamefont {Golden}}, \bibinfo {author}
  {\bibfnamefont {H.}~\bibnamefont {Seki}}, \bibinfo {author} {\bibfnamefont
  {C.~A.}\ \bibnamefont {Brown}}, \ and\ \bibinfo {author} {\bibfnamefont
  {M.~S.}\ \bibnamefont {de~Vries}},\ }\href@noop {} {\bibfield  {journal}
  {\bibinfo  {journal} {Chem. Phys. Lett.},\ }\textbf {\bibinfo {volume}
  {179}},\ \bibinfo {pages} {181} (\bibinfo {year} {1991})}\BibitemShut
  {NoStop}%
\bibitem [{\citenamefont {Hedin}\ and\ \citenamefont
  {Lundqvist}(1969)}]{Hedin1969a}%
  \BibitemOpen
  \bibfield  {author} {\bibinfo {author} {\bibfnamefont {L.}~\bibnamefont
  {Hedin}}\ and\ \bibinfo {author} {\bibfnamefont {S.}~\bibnamefont
  {Lundqvist}},\ }in\ \href@noop {} {\emph {\bibinfo {booktitle} {Solid State
  Physics}}},\ Vol.~\bibinfo {volume} {23},\ \bibinfo {editor} {edited by\
  \bibinfo {editor} {\bibfnamefont {H.}~\bibnamefont {Ehrenreich}}, \bibinfo
  {editor} {\bibfnamefont {D.}~\bibnamefont {Turnbull}}, \ and\ \bibinfo
  {editor} {\bibfnamefont {F.}~\bibnamefont {Seitz}}}\ (\bibinfo  {publisher}
  {Academic Press},\ \bibinfo {address} {New York},\ \bibinfo {year} {1969})\
  p.~\bibinfo {pages} {1}\BibitemShut {NoStop}%
\bibitem [{\citenamefont {Child}\ and\ \citenamefont
  {Longuet-Higgins}(1961)}]{Child1961a}%
  \BibitemOpen
  \bibfield  {author} {\bibinfo {author} {\bibfnamefont {M.~S.}\ \bibnamefont
  {Child}}\ and\ \bibinfo {author} {\bibfnamefont {H.~G.}\ \bibnamefont
  {Longuet-Higgins}},\ }\href@noop {} {\bibfield  {journal} {\bibinfo
  {journal} {Phil. Trans. R. Soc. A},\ }\textbf {\bibinfo {volume} {254}},\
  \bibinfo {pages} {259} (\bibinfo {year} {1961})}\BibitemShut {NoStop}%
\bibitem [{\citenamefont {Feynman}(1939)}]{Feynman1939a}%
  \BibitemOpen
  \bibfield  {author} {\bibinfo {author} {\bibfnamefont {R.~P.}\ \bibnamefont
  {Feynman}},\ }\href@noop {} {\bibfield  {journal} {\bibinfo  {journal} {Phys.
  Rev.},\ }\textbf {\bibinfo {volume} {56}},\ \bibinfo {pages} {340} (\bibinfo
  {year} {1939})}\BibitemShut {NoStop}%
\bibitem [{\citenamefont {Becke}(1993)}]{Becke1993a}%
  \BibitemOpen
  \bibfield  {author} {\bibinfo {author} {\bibfnamefont {A.~D.}\ \bibnamefont
  {Becke}},\ }\href@noop {} {\bibfield  {journal} {\bibinfo  {journal} {J.
  Chem. Phys.},\ }\textbf {\bibinfo {volume} {98}},\ \bibinfo {pages} {5648}
  (\bibinfo {year} {1993})}\BibitemShut {NoStop}%
\bibitem [{\citenamefont {Sato}\ \emph {et~al.}(2006)\citenamefont {Sato},
  \citenamefont {Tokunaga},\ and\ \citenamefont {Tanaka}}]{Sato2006a}%
  \BibitemOpen
  \bibfield  {author} {\bibinfo {author} {\bibfnamefont {T.}~\bibnamefont
  {Sato}}, \bibinfo {author} {\bibfnamefont {K.}~\bibnamefont {Tokunaga}}, \
  and\ \bibinfo {author} {\bibfnamefont {K.}~\bibnamefont {Tanaka}},\
  }\href@noop {} {\bibfield  {journal} {\bibinfo  {journal} {J. Chem. Phys.},\
  }\textbf {\bibinfo {volume} {124}},\ \bibinfo {pages} {024314} (\bibinfo
  {year} {2006})}\BibitemShut {NoStop}%
\bibitem [{\citenamefont {Sato}\ \emph {et~al.}(2009)\citenamefont {Sato},
  \citenamefont {Tokunaga}, \citenamefont {Iwahara}, \citenamefont {Shizu},\
  and\ \citenamefont {Tanaka}}]{Sato2009a}%
  \BibitemOpen
  \bibfield  {author} {\bibinfo {author} {\bibfnamefont {T.}~\bibnamefont
  {Sato}}, \bibinfo {author} {\bibfnamefont {K.}~\bibnamefont {Tokunaga}},
  \bibinfo {author} {\bibfnamefont {N.}~\bibnamefont {Iwahara}}, \bibinfo
  {author} {\bibfnamefont {K.}~\bibnamefont {Shizu}}, \ and\ \bibinfo {author}
  {\bibfnamefont {K.}~\bibnamefont {Tanaka}},\ }in\ \href@noop {} {\emph
  {\bibinfo {booktitle} {The Jahn--Teller Effect: Fundamentals and Implications
  for Physics and Chemistry}}},\ \bibinfo {editor} {edited by\ \bibinfo
  {editor} {\bibfnamefont {H.}~\bibnamefont {K{\"{o}}ppel}}, \bibinfo {editor}
  {\bibfnamefont {D.~R.}\ \bibnamefont {Yarkony}}, \ and\ \bibinfo {editor}
  {\bibfnamefont {H.}~\bibnamefont {Barentzen}}}\ (\bibinfo  {publisher}
  {Springer--Verlag},\ \bibinfo {address} {Berlin and Heidelberg},\ \bibinfo
  {year} {2009})\ p.~\bibinfo {pages} {99}\BibitemShut {NoStop}%
\bibitem [{\citenamefont {Frisch}\ \emph {et~al.}(2004)\citenamefont {Frisch}
  \emph {et~al.}}]{g03}%
  \BibitemOpen
  \bibfield  {author} {\bibinfo {author} {\bibfnamefont {M.~J.}\ \bibnamefont
  {Frisch}} \emph {et~al.},\ }\href@noop {} {\emph {\bibinfo {title} {Gaussian
  03, Revision E.01}}},\ \bibinfo {address} {Wallingford, CT} (\bibinfo {year}
  {2004})\BibitemShut {NoStop}%
\bibitem [{\citenamefont {Narymbetov}\ \emph {et~al.}(1999)\citenamefont
  {Narymbetov}, \citenamefont {Kobayashi}, \citenamefont {Tokumoto},
  \citenamefont {Omerzu},\ and\ \citenamefont {Mihailovic}}]{Narymbetov1999a}%
  \BibitemOpen
  \bibfield  {author} {\bibinfo {author} {\bibfnamefont {B.}~\bibnamefont
  {Narymbetov}}, \bibinfo {author} {\bibfnamefont {H.}~\bibnamefont
  {Kobayashi}}, \bibinfo {author} {\bibfnamefont {M.}~\bibnamefont {Tokumoto}},
  \bibinfo {author} {\bibfnamefont {A.}~\bibnamefont {Omerzu}}, \ and\ \bibinfo
  {author} {\bibfnamefont {D.}~\bibnamefont {Mihailovic}},\ }\href@noop {}
  {\bibfield  {journal} {\bibinfo  {journal} {Chem. Commun.},\ \bibinfo {pages}
  {1511}} (\bibinfo {year} {1999})}\BibitemShut {NoStop}%
\bibitem [{\citenamefont {Fujiwara}\ \emph {et~al.}(2005)\citenamefont
  {Fujiwara}, \citenamefont {Kambe},\ and\ \citenamefont
  {Oshima}}]{Fujiwara2005a}%
  \BibitemOpen
  \bibfield  {author} {\bibinfo {author} {\bibfnamefont {M.}~\bibnamefont
  {Fujiwara}}, \bibinfo {author} {\bibfnamefont {T.}~\bibnamefont {Kambe}}, \
  and\ \bibinfo {author} {\bibfnamefont {K.}~\bibnamefont {Oshima}},\
  }\href@noop {} {\bibfield  {journal} {\bibinfo  {journal} {Phys. Rev. B},\
  }\textbf {\bibinfo {volume} {71}},\ \bibinfo {pages} {174424} (\bibinfo
  {year} {2005})}\BibitemShut {NoStop}%
\bibitem [{\citenamefont {David}\ \emph {et~al.}(1991)\citenamefont {David},
  \citenamefont {Ibberson}, \citenamefont {Matthewman}, \citenamefont
  {Prassides}, \citenamefont {Dennis}, \citenamefont {Hare}, \citenamefont
  {Kroto}, \citenamefont {Tayloa},\ and\ \citenamefont {Walton}}]{David1991a}%
  \BibitemOpen
  \bibfield  {author} {\bibinfo {author} {\bibfnamefont {W.~I.~F.}\
  \bibnamefont {David}}, \bibinfo {author} {\bibfnamefont {R.~M.}\ \bibnamefont
  {Ibberson}}, \bibinfo {author} {\bibfnamefont {J.~C.}\ \bibnamefont
  {Matthewman}}, \bibinfo {author} {\bibfnamefont {K.}~\bibnamefont
  {Prassides}}, \bibinfo {author} {\bibfnamefont {T.~J.~S.}\ \bibnamefont
  {Dennis}}, \bibinfo {author} {\bibfnamefont {J.~P.}\ \bibnamefont {Hare}},
  \bibinfo {author} {\bibfnamefont {H.~W.}\ \bibnamefont {Kroto}}, \bibinfo
  {author} {\bibfnamefont {R.}~\bibnamefont {Tayloa}}, \ and\ \bibinfo {author}
  {\bibfnamefont {D.~R.~M.}\ \bibnamefont {Walton}},\ }\href@noop {} {\bibfield
   {journal} {\bibinfo  {journal} {Nature (London)},\ }\textbf {\bibinfo
  {volume} {353}},\ \bibinfo {pages} {147} (\bibinfo {year}
  {1991})}\BibitemShut {NoStop}%
\bibitem [{\citenamefont {B{\"{u}}rgi}\ \emph {et~al.}(1992)\citenamefont
  {B{\"{u}}rgi}, \citenamefont {Blanc}, \citenamefont {Schwarzenbach},
  \citenamefont {Liu}, \citenamefont {Lu},\ and\ \citenamefont
  {Kappes}}]{Burgi1992a}%
  \BibitemOpen
  \bibfield  {author} {\bibinfo {author} {\bibfnamefont {H.~B.}\ \bibnamefont
  {B{\"{u}}rgi}}, \bibinfo {author} {\bibfnamefont {E.}~\bibnamefont {Blanc}},
  \bibinfo {author} {\bibfnamefont {D.}~\bibnamefont {Schwarzenbach}}, \bibinfo
  {author} {\bibfnamefont {S.~Z.}\ \bibnamefont {Liu}}, \bibinfo {author}
  {\bibfnamefont {Y.~J.}\ \bibnamefont {Lu}}, \ and\ \bibinfo {author}
  {\bibfnamefont {M.}~\bibnamefont {Kappes}},\ }\href@noop {} {\bibfield
  {journal} {\bibinfo  {journal} {Angew. Chem. Int. Ed. Engl.},\ }\textbf
  {\bibinfo {volume} {31}},\ \bibinfo {pages} {640} (\bibinfo {year}
  {1992})}\BibitemShut {NoStop}%
\bibitem [{0-0()}]{0-0}%
  \BibitemOpen
  \href@noop {} {}\bibinfo {note} {Actually this line contains contributions
  also from temperature populated excited vibronic levels of C$_{60}^-$ to some
  excited vibrational levels of C$_{60}$. However in the present case their
  contribution to the 0--0 line is very small.}\BibitemShut {Stop}%
{
\bibitem [{\citenamefont {Bruyndonckx}\ \emph {et~al.}(1997)\citenamefont
  {Bruyndonckx}, \citenamefont {Daul}, \citenamefont {Manoharan},\ and\
  \citenamefont {Deiss}}]{Bruyndonckx1997a}%
  \BibitemOpen
  \bibfield  {author} {\bibinfo {author} {\bibfnamefont {R.}~\bibnamefont
  {Bruyndonckx}}, \bibinfo {author} {\bibfnamefont {C.}~\bibnamefont {Daul}},
  \bibinfo {author} {\bibfnamefont {P.~T.}\ \bibnamefont {Manoharan}}, \ and\
  \bibinfo {author} {\bibfnamefont {E.}~\bibnamefont {Deiss}},\ }\href@noop {}
  {\bibfield  {journal} {\bibinfo  {journal} {Inorg. Chem.},\ }\textbf
  {\bibinfo {volume} {36}},\ \bibinfo {pages} {4251} (\bibinfo {year}
  {1997})}\BibitemShut {NoStop}%
}
\end{thebibliography}

%

\end{document}